\documentclass[onecolumn]{IEEEtran}
\usepackage{graphicx}
\usepackage{caption}
\usepackage{hyperref}
\usepackage{amssymb}
\usepackage{amsmath}

\begin{document}

\title{A Review of Theoretical and Practical Challenges of Trusted Autonomy in Big Data}

\author{Hussein A. Abbass, George Leu, and Kathryn Merrick
 \thanks{Authors are with the School of Engineering \& IT, University of New South Wales,
 Canberra-Australia, (e-mail: \{h.abbass,g.leu,k.merrick\}@adfa.edu.au).}
}

%



\maketitle

\begin{abstract}
Despite the advances made in artificial intelligence, software
agents, and robotics, there is little we see today that we can
truly call a fully autonomous system. We conjecture that the main
inhibitor for advancing autonomy is lack of trust. Trusted
autonomy is the scientific and engineering field to establish the
foundations and ground work for developing trusted autonomous
systems (robotics and software agents) that can be used in our
daily life, and can be integrated with humans seamlessly,
naturally and efficiently.

In this paper, we review this literature to reveal opportunities
for researchers and practitioners to work on topics that can
create a leap forward in advancing the field of trusted autonomy.
We focus the paper on the `trust' component as the uniting
technology between humans and machines. Our inquiry into this
topic revolves around three sub-topics: (1) reviewing and
positioning the trust modelling literature for the purpose of
trusted autonomy; (2) reviewing a critical subset of sensor
technologies that allow a machine to sense human states; and (3)
distilling some critical questions for advancing the field of
trusted autonomy. The inquiry is augmented with conceptual models
that we propose along the way by recompiling and reshaping the
literature into forms that enables trusted autonomous systems to
become a reality. The paper offers a vision for a Trusted Cyborg
Swarm, an extension of our previous Cognitive Cyber Symbiosis
concept, whereby humans and machines meld together in a
harmonious, seamless, and coordinated manner.
\end{abstract}


 \section{Introduction}


Trusted Autonomous Systems (TAS) are one of the grand challenges
in this century. The proliferation of autonomous systems (AS) is a
natural consequence of the opportunities they create for industry,
health, education and the economy as a whole. These opportunities
arrive with new technological and social challenges. TAS can be
seen as a fusion of these challenges and acts as the umbrella that
couples underneath it both objectives and challenges concurrently.
Trusted Autonomy is the wider scientific endeavor to establish the
science and engineering required to develop TAS.

For most of the literature, trust is seen as a human endeavor. It
is seen to be owned by fields such as
psychology~\cite{delgado2005perceptions11,deutsch1977resolution13},
social
sciences~\cite{deutsch1962cooperation12,deutsch1977resolution13,luhmann1979trust32},
and organization
sciences~\cite{wells2001trust43,reina2000trust23,whitener1998managers44,spector2004trust41}.
Technologists and computational scientists have been modelling
trust for a few
decades~\cite{Marsh1994,Pinyol2013,petraki2014trust,abbass2015trusted}.
TAS brings this large spectrum of research together in the face of
a new challenge: the creation of mutual trust between a human and
a machine.

The above challenge carries forward classic research in human
factors, automation, human computer interaction, and human robotic
interaction to tackle the problem of how to get a human to trust a
machine. However, in addition to this one-sided view of trust, the
challenge is raising a more fundamental complex question on the
other side of the equation: how can a machine trust a human?

This latter question is a socio-technological question on
modelling complex systems, and forms the context of this paper.
For a machine to trust a human, it needs to be able to
`proactively' sense the human, extract meaning and cues during
sensing, manage identity to ensure it is dealing with the right
human, estimate and manage human intent, and make an informed
judgement on the level of trustworthiness of the human in a
particular context in time and space.

Finding technological solutions to address this question is a
complex endeavor that is compounded even more with the dynamic and
possible chaotic nature of a trusting relationship. The slightest
change in context may cause a dramatic change in trust estimation
with a profound impact on trust dynamics. For example, a slight
change in the security level of the communication channel between
the human and the machine can cause a dramatic change in trust
estimation.

To put it more succinctly, TAS require the machine to possess a
high level of intelligence extending beyond the level of the task
it is performing to include the level of the interaction it is
involved with, the environment it is `situated' and `embodied'
within, and to be aware of its own states, actions, motives,
goals, and action production mechanisms. TAS require smart
technologies for the machine to manage the interaction with
humans; not just to convince the human that the machine is
trustworthy, but to be `self-aware' of trust to the extent of
making its own decisions on the varying degree of trustworthiness
of the human operators/users themselves during the interaction and
based on context.

Designing the Artificial Intelligence (AI) models to manage the
interaction between the human and the machine in the way explained
above requires some fundamental building blocks. We structure this
paper into three main sections.

The first section (Section~\ref{CFBT}) focuses on trust as the
fundamental differentiator between classic AS research and TAS. We
first start with the layers needed to define trust, then contrast
how trust is modelled in the literature with these layers to offer
two conceptual models: one is a computational structure model of
trust and the other is a trust-based multi-agent architecture. We
then define the `trust aware' agent and present the concept of
`Trust Bus', which carries the information resultant from action
production within the agent to the actuators. we adopt the term
'bus' in its computer architecture sense as in its classic use in
computer architecture and service oriented architectures. The lens
of the Trust Bus scopes the big data problem within this context
to particular set of data and sensors. This scoping is essential
as it guides the selection of sensors to be reviewed in the next
section.

The second section (Section~\ref{BDHMI}) focuses on introducing
big data to emphasize that the word `big' is a relative concept
defined in relation to the amount of processing available. We then
review the sensor technologies available today that can facilitate
the flow of information between humans and machines and augment
the review with an overview of different processing concepts
available to process these information. We conclude this section
with a discussion on why the human-machine relationship is a big
data problem.

In the third section (Section~\ref{TASBDCR}), we extend the
discussion to the concept of autonomy and discuss the layers of
autonomy. We then conclude this section by bringing together the
first two sections and, first, offer the architecture required for
an autonomous entity to exhibit both autonomy and trust, then
second discuss the challenges associated with TAS from big data
perspective.

 \section{Computations for and by Trust}\label{CFBT}

 \subsection{The Road to Trust}

Trust is a multi-dimensional concept. Research has established
many facets and influential
factors~\cite{abbass2015trusted,petraki2014trust} for trust
dynamics~\cite{abbass2015n}. We will discuss these facets and
factors incrementally as each factor requires the list of factors
listed before it to have a meaningful evolution of a trusting
relationship.

 \begin{figure*}[!t]
 \centering
 \captionsetup{justification=centering}
 \includegraphics[width=12cm,height=4cm]{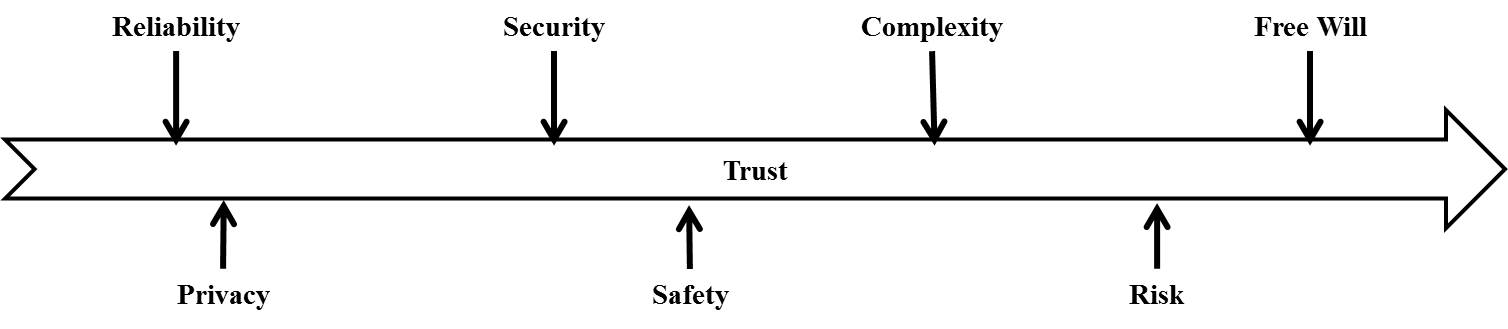}
 \caption{Trust Timeline.} \label{TrustTimeline}
 \end{figure*}

The evolution model of trust presented in
Figure~\ref{TrustTimeline} starts with the concept of reliability.
A trustor requires the trustee to be reliable; that is, the
trustee needs to be consistent in its actions, values, and
objectives during the lifetime of the relationship and stable in
its performance. Fundamentally, a trusting relationship carries
with it a level of risk; that is, the risk that the trustee will
default. Reliability increases the confidence level associated
with predictability. Therefore, it offers a better estimate of the
risk a trustor is exposed to during the trusting relationship.

Once reliability is established, trust improves through privacy.
The order is a very important factor in this model. In the absence
of any contextual information, a trustee that maintains privacy
but not reliability is a high risk trustee. The reverse is not
necessarily true as it depends on context. The trustor requires
assurance that private information in the trusting relationship
are protected.

Privacy is a distinct concept from security. When a trustor trusts
a trustee, the trusting relationship involves transfer of certain
data. For example, the trustee may need to know the identity of
the trustor and vice-versa. These data are communicated between
parties for the purpose of fulfilling the trusting relationship.
Any misuse of these data outside terms of the trusting contract is
a breach of privacy.

Security, however, is more related to what is known as the CIA
triad~\cite{alruwaili2015secsdlc}: Confidentiality, Integrity and
Availability. Confidentiality carries an element of privacy. A
security breach where the database held by the trustee is broken
into by an unauthorized third part can also be a breach of
privacy. Maybe the third party only stole the trustee's data
because it is not interested in the trustor's data per se; thus,
the trustor data may not have been compromised. In this case,
there is no privacy breach from the trustor perspective but a
clear security breach that may compromise the trustor confidence
in the trusting relationship.

Integrity breaches can be unintentional. A simple error in a piece
of code can cause a deletion or modification of data and
compromising the accuracy and validity of information.
Availability is a function of the reliability of the IT service
such that access is guaranteed whenever it is required and access
control such that only authorized actors can access the data.

The above discussion can be seen from a reliability lens, where
the concept of reliability discussed above is focused on
reliability for delivering the service while the privacy and
security elements relate to the reliability of the information
assurance process.

The fourth factor is related to safety~\cite{hinchman2012towards}.
A trustor needs to be safe in a trusting relationship. In a safety
critical system, like air traffic management, the air traffic
controller issues commands to pilots to ensure that the aircraft
are separated (separation assurance). The pilot plays the trustor
since the pilot does not have access to all information related to
the airspace and needs to trust the air traffic controller
judgement. The air traffic controller is the trustee. The trustee
is highly trained and reliable. The communication channel and data
within this safety critical system are subject to strict
information assurance procedures. But what will happen if the
working environment for the air traffic controller is not safe?
For example, there is a high possibility for a fire hazard, loss
of power, etc. In this case, safety, or lack of in this example,
can alter reliability of service and information assurance.

The four factors discussed so far are related to what we will
call, technological imperfection. They are imperfect because we
can't guarantee a 100\% reliability, privacy, security and safety
in any situation. Regardless of how much we will invest, the laws
of dimensioning return would prevail.

The remaining three factors are related to what we will call
`cognitive imperfection' and start coming into play. The first
factor is related to the complexity of the context for the trustor
and trustee. A trustor would normally enter a trusting
relationship to manage some level of complexity. Trust is a form
of educated delegation. A trustor would delegate to a trustee when
the trustor `needs to'; that is, there is a benefit for the
trustor in trusting than self-performing the job. This delegation
needs to reduce some form of complexity, be it, the technical
complexity associated with performing the task, the time pressure
on the trustor to complete the task, or the increase in mental and
cognitive complexity on the trustor if the trustor chooses to
perform the task. As the level of complexity increases, the degree
with which a trustor trusts a trustee increases.

The second factor of cognitive imperfection is related to risk. As
we mentioned before, a trusting decision involves a level of
uncertainty associated with the possibility that the trustee
defects~\cite{petraki2014trust,abbass2015n}. Risk within a
trusting decision is both negative and positive. The expectations
from the trustor towards the good behavior of the trustee is a
positive risk. It carries with it the negative risk resultant from
the trustee defecting in the relationship. Taking high risk is an
indication of high trust, while the reverse is not necessarily
true.

The last factor in the evolutionary timeline of trust is
`free-will'; the fact that the trustor is making the trusting
decision not because of a necessity forced by an external actor
who is attempting to exert its influence on the trustor. With
free-will arises full autonomy. An actor that is fully autonomous
acts because it has the freedom to act. This freedom is not
unbounded since social ties, social rules and norms, and
interdependencies among actors in terms of resources and
objectives limit the behavior of an actor. Free-will is the
ability of the actor to make a decision within this bounded space
autonomously and at its own discretion. This discussion begs the
question of what is autonomy?

 \subsection{Computational Trust}

In multi-agent systems, agents can cooperate to achieve goals that
may not be attainable by individual agents. However, agents must
manage the risk associated with interacting with others who may
have different objectives, or who may fail to fulfil their
commitments during a cooperative scenario. Models of trust offer a
mechanism for reasoning about the reliability, honesty, veracity
etc. of other agents, to provide a basis for determining whether
cooperation is worthwhile. This section is a selective survey of
computational models of trust that have been used in multi-agent
systems, including the types of trust models, the agent-models in
which they have been used, the type of cooperation achieved and
the applications in which trusting agents have been used.

A number of existing surveys of computational trust have been made
\cite{Sabater2005,Pinyol2013,Josang2007,Yu2013}. Sabater and
Sierra \cite{Sabater2005} classify trust and reputation models
firstly as either cognitive or game theoretic. They also consider
the information sources for trust, including direct experiences,
indirect witness information and sociological information.

This section also considers these dimensions of trust models, but
goes beyond this to survey the types of agent models in which
trust models have been embedded, the applications for which they
have been used and techniques for evaluating trust models. We also
look beyond the scope of traditional agent-based systems
literature to consider trust models from mathematical
optimisation. The rest of this subsection is divided into two
parts: trust modelling and computations by trust.

 \subsubsection{Trust Modelling}

Computational models of trust are essentially mathematical
functions that aim to represent certain aspects of philosophical
or psychological theories of trust. The earliest computational
models date back to the 1990s and are relatively simple
statistical functions taking into account factors such as the
situation of an agent, the utility of the situation and the
importance of the situation. Later models have emerged in time
with the development of more complex machine learning paradigms.

We first take the approach of Josang et al. \cite{Josang2007},
dividing the basic trust models into categories for statistical
models, Bayesian analysis, discrete models, belief models, fuzzy
models and flow models. However, we also extend this set of
categories with a discussion of trust-regions from the literature
of mathematical optimisation and computer networks.

\begin{itemize}

\item {\bf Statistical Models:} An early approach to modelling
trust was proposed by Marsh \cite{Marsh1994a}. Marsh presents
several definitions of trust that take into account different
factors. First, he defines ``the amount $x$ trusts $y$'' by
$T_{x}(y)$. $T_{x}(y)$ has a value in the interval $[-1;1)$ (i.e.,
$-1 \leq T_{x}(y)<1$). In his model, $0$ means no trust, and $-1$
represents total distrust. He argues that the two are not the same
since, the situation of `no trust' occurs when the trustor has
little or no information about the trustee, or is indifferent. In
contrast, `total distrust' is a proactive measure, requiring that
the trustor reason about what it is doing. Marsh does not permit a
value of 1 for trust, arguing that `blind trust' implies that the
trustor does not reason about the situation, and this violates
philosophical definitions of trust. In a second definition, Marsh
includes the notion of a situation $\alpha$. A situation is a
point or points in time relative to a specific agent. The
importance of the situation to an agent $x$ is written
$I_{x}(\alpha)$. $I_{x}(\alpha)$ has a value in $(-1, 1)$. The
utility (cost or benefit) of a situation to an agent $x$ is
written $U_{x}(\alpha)$ also with a value in $(-1, 1)$. The
``situational trust'' of agent $x$ for agent $y$ in situation
$\alpha$ is then defined as:
\begin{equation}
    T_{x}(y,\alpha)=T_{x}(y)U_{x}(\alpha)I_{x}(\alpha)
    \label{eq:Trust}
\end{equation}

Estimates for each of these terms must then be obtained. Marsh
assumes that values of $U_{x}(\alpha)$ and $I_{x}(\alpha)$ can be
determined from the domain. Marsh suggests that the amount $x$
trusts $y$ at time $t$, might be determined as an average of
situational trust values at previous times. Because agent memory
is assumed to be bounded, one challenge is to decide which
situations should be included in this average. Alternatives
include all the situations the agent remembers or only situations
that were similar to the current situation and so on. Different
choices will clearly result in different trust values.

Other models have built on the work of Marsh, and maintain the
idea of trust as a scalar value. Griffiths and Luck
\cite{Griffiths2003} define the trust in an agent $y$, to be a
value from the interval between $0$ and $1$: $T_{y}\in[0, 1]$.
These numbers represent comparative values, and are not considered
meaningful in themselves. Values approaching $0$ are defined to
represent complete distrust, and values approaching $1$ represent
complete or `blind' trust. This is thus a simplified model,
compared to that proposed by Marsh.

\item \textbf{Bayesian Analysis:} More recent models take a
Bayesian approach to modelling trust \cite{LukeTeacy2006}. Rather
than using `situations' as the basis for calculation, binary
valued interactions form the input for the trust calculation. An
interaction $O_{x,y}$ takes a value of $1$ if $y$ fulfilled a
contract with $x$, and $0$ otherwise. `Interactions' by this
definition thus have a more constrained structure compared to
Marsh's situations. The tendency of an agent y to fulfil or
default on its obligations is governed by its behaviour, which is
represented as a variable $B_{x,y}\in[0,1]$. $B_{x,y}$ specifies
the intrinsic probability that $x$ will fulfil its obligations
during an interaction with $y$. That is, $B_{x,y}=Pr(O_{x,y}=1)$.
The level of trust $T_{x,y}$ is defined as the expected value of
$B_{x,y}$ given a set of $K$ outcomes $O_{x,y}^{1:K}$. That is:
\begin{equation}
T_x{y}=E(B_{x,y}|O_{x,y}^{1:K}) \label{eq:BATrust}
\end{equation}

In order to determine this expected value, a probability
distribution defined by a probability density function (PDF) is
used to model the relative probability that $B_{x,y}$ will have a
certain value. An example is a beta PDF (see
Equation~\ref{eq:Collective} for example).

\item \textbf{Discrete Models:} Other models stipulate a finite,
discrete set of trust categories
\cite{Abdul-Rahman2000,Cahill2003}. For example, the model
proposed by Abdul-Rahman and Hailes \cite{Abdul-Rahman2000}
stipulates four categories: very trustworthy, trustworthy,
untrustworthy and very untrustworthy. An agent's trust degree is
represented as a tuple, rather than a scalar value:
\begin{equation}
T=\{vt,t,u,vu\} \label{eq:DMTrust}
\end{equation}
Trust is equal to the trust degree with the highest value. One
agent's $x$ belief in another agent $y$'s trustworthiness in a
given context $\alpha$ is represented as:
\begin{equation}
    belief_x(y,\alpha,T)
    \label{eq:Belief}
\end{equation}
An agent maintains this information for each other agent and
context.

\item \textbf{Belief Models:} A variation on discrete models is
belief-based model such as those proposed by Josang
\cite{Josang1999}. In their model, trust is modelled as an opinion
held by $x$ about another agent $y$:
\begin{equation}
\omega_y^x=(b,d,u,a), \ \ \ b, d, u, a\in[0, 1] \label{eq:Opinion}
\end{equation}
where $b$ represents belief, $d$ disbelief and $u$ uncertainty.
$a$ is the base-rate probability in the absence of evidence.

\item \textbf{Fuzzy Models:} Fuzzy models provide another way to
deal with the uncertainty associated with trust
\cite{Griffiths2006,Aref2014}. A fuzzy set is a pair $(X,m)$ where
$X$ is a set and $m:X\to[0, 1]$ is a membership function that
specifies the grade of membership of $x$ in $(X,m)$. For each $x
\in X$, then $x$ is not included in $(X,m)$ if $m(x)=0$.
Conversely it is fully included in $(X,m)$ if $m(x)=1$. If we
maintain our definition of $x$ as an agent, then a fuzzy set
$(X,m)$ may be defined as very trusted agents, trusted agents,
untrusted agents and so on.

\item \textbf{Flow Models:} Josang et al. \cite{Josang2007} define
flow models as those that compute trust by transitive iteration
through looped or arbitrarily long chains. The Advogato maximum
flow trust metric \cite{Levien2009}, for example, computes trust
of individuals in a group $V$ relative to a `seed' $s$ of highly
trusted group members. The input of the trust calculation are an
integer number $n$, which represents the number of group members
to trust, as well as the trust seed $s$. The output is a
characteristic function that maps each member to a boolean value
indicating trustworthiness:

\begin{equation}
T : 2^V \times \mathbb{N}_0^+ \to (V \to \{ true,false \})
\label{eq:Fuzzy}
\end{equation}

The trust model underlying Advogato does not provide support for
weighted trust relationships. Hence, trust edges extending from
individual $x$ to $y$ express blind trust of $x$ in $y$. Other
examples of flow based metrics include the Appleseed trust metric
\cite{Ziegler2004}.

A characteristic component of flow models is the seed (also called
the trust root). The seeds of trust, are the positive assumptions
about specific entities made by all entities in some community.
The label 'authority' is often applied to an entity that is
the subject of a trust root. Models with formally modelled trust
seeds may be termed centralized and those without may be termed
distributed \cite{Mahoney2005}.

\item \textbf{Optimisation Models:} Another approach to modelling
trust has been used in mathematical optimisation \cite{Conn2000}.
This this field, a `trust region' denotes the part of an objective
function for which there is a good approximate model function. An
example of a model function here is a quadratic surface.
Algorithms that incorporate trust region methods are known as
restricted step methods, and broadly take the following approach
\cite{Yuan1999}: If a `good' model of the objective function is
found within the trust region, then the region is expanded.
Conversely, if the model is poor, then the region is contracted.
The quality of the model is evaluated by comparing the ratio of
expected improvement from the model approximation with the actual
improvement observed in the objective function. Thresholding of
this ratio can be used as the criterion for expansion and
contraction of the trust region. The model function is then
'trusted' only in the region where it provides a good
approximation.

\end{itemize}

The models discussed above assume that trust is the same for all
agents, essentially an objective model of trust. However,
alternative models, such as that proposed by Marsh
\cite{Marsh1994a}, discuss the role of the disposition of the
trustor in computing trust. This interprets trust as a subjective
concept. Marsh presents a spectrum of dispositions from optimists
to pessimists. Optimists are more likely to trust, while
pessimists are less likely to trust. In contrast to the generic,
average-based definition of trust described above, he defines
optimists to compute trust as in Equation~\ref{eq:Optimist}, while
pessimists compute trust as in Equation~\ref{eq:Pesimist}:

\begin{equation}
T_x(y)=max_{\alpha \in A} (T_x(y,\alpha)) \label{eq:Optimist}
\end{equation}

\begin{equation}
T_x(y)=min_{\alpha \in A} (T_x(y,\alpha)) \label{eq:Pesimist}
\end{equation}

Again, the choice of which situations to include will influence
the value of trust. Griffiths and Luck \cite{Griffiths2003} use
dispositions as a way of initialising trust values. In their
model, trust values are initially inferred according to an agent's
disposition: optimistic agents infer high values, while pessimists
infer low values. Disposition also assumed to determine how trust
is updated after interactions in their model.

An assumption of the models above is that trust values depend on
factors such as the agent's disposition and its personal
experiences with the trustee. When assessing trust in an
individual with whom an agent has no personal experience,
information must be obtained in another way. One method is to
consult 'witnesses' to develop a collective opinion. This
collective opinion is often referred to as 'reputation', and can
be used as a factor in the assessment of trustworthiness. The
following is a short review of computational models of trust that
take this approach:

\begin{itemize}

\item \textbf{Statistical Models:} A simple model permits
witnesses to report on a transaction by assigning one of three
possible values $1$ (positive), $0$ neutral or $-1$ (negative).
Reputation is computed simply as the sum or an average of all
reported values over a certain time period \cite{Sabater2005}.
More complex variants of this approach weight recent witness
reports more highly than older reports, or stipulate different
rates of reputation change depending on the current reputation of
an agent \cite{Giorgos1999}.

\item \textbf{Bayesian Analysis:} Still more complex variants take
into account specific details of interactions between individuals
to compute reputation. One such model \cite{LukeTeacy2006}
represents $x$'s opinion of $y$'s reputation as:
\begin{equation}
    R_{x,y}^t=(m_{x,y}^t,n_{x,y}^t)
    \label{eq:Reputation}
\end{equation}
where $m_{x,y}^t$ is the number of successful interactions
(interactions where $O_{x,y}=1$) and $n_{x,y}^t$ is the number of
unsuccessful interactions (interactions where $O_{x,y}=0$)
experienced by $x$. $x$ may communicate this opinion with a third
agent $z$. The communicated opinion $\hat{R}̂_{x,y}^t$ may be
distinguished from the true opinion $R_{x,y}^t$ as is possible
that $x$ may not report accurately for a range of reasons. $z$ can
develop a collective opinion of $y$ by summing values of
$\hat{m}_{x,y}^t$ and $\hat{n}_{x,y}^t$ collected from multiple
other agents $x$ and using these values to calculate the shape
parameters for a beta distribution, as in
Equation~\ref{eq:Collective}.
\begin{equation}
    \begin{split}
        & N_{z,y}=\sum_x \hat{n}_{x,y} ~~ and ~~ M_{z,y}=\sum_x \hat{m}_{x,y} \\
        & \alpha = M_{z,y}+1 ~~ and ~~ \beta = N_{z,y}+1 \\
        & f(B_{z,y}|\alpha,\beta) = \frac {(B_{z,y})^{\alpha-1}(1-B_{z,y})^{\beta-1}}
        {\int\limits_0^1 u^{\alpha-1}(1-u)^{\beta-1}\mathrm{d}u}
    \end{split}
    \label{eq:Collective}
\end{equation}
The final value for trust is calculated by applying the standard
equation for the expected value of a beta distribution to these
parameter settings: $T_{z,y}^t = E(B_{z,y}|\alpha,\beta) =
\frac{\alpha}{\alpha + \beta}$. Then, $z$ may also include their
own opinion in the trust calculation if such data exists. If all
agents have complete information about the true opinions of all
other agents, then it is said that a global or objective measure
of trust results. Otherwise, trust remains local or subjective
\cite{Ziegler2004}.

\item \textbf{Network Models:} Another set of trust models based
on witness data have been proposed for use in computer networks,
and peer-to-peer computing, particularly in applications where
multi-agent systems are responsible for transporting valuable data
over a network. Examples of trust models for computer networks
include PeerTrust \cite{Xiong2004}, FCTrust \cite{Hu2008} and
SFTrust \cite{Zhang2009}. Other models build on this work
\cite{Xiong2004}. They use parameters such as the feedback a peer
obtains from other peers (also called satisfaction); feedback
scope (such as the total number of transactions a peer has with
other peers); the credibility factor for the feedback source; the
transaction context factor for discriminating critical from less
critical transactions; and the community context factor for
addressing community-related characteristics and vulnerabilities.
An example of a network trust metric based on these parameters is:
\begin{equation}
    T(x)=\alpha \sum_{i=1}^{I(x)} S(x,i)Cr[p(x,i)]TF(x,i)+ \beta CF(x) \label{eq:NetTrust}
\end{equation}
$I(x)$ denotes the total number of transactions performed by peer
$x$ with all other peers. $p(x,i)$ denotes the other participating
peer in peer $x$'s $i^{th}$ transaction. $S(x,i)$ denotes the
normalised satisfaction peer $x$ receives from peer $p(x,i)$ in
its $i^{th}$ transaction. $Cr(.)$ denotes the credibility of
feedback, $TF(x,i)$ denotes the adaptive transaction context
factor for peer $x$'s $i^{th}$ transaction. $CF(x)$ denotes the
community context factor. $\alpha$ and $\beta$ denote the
normalise weight factors for the collective evaluation and the
community context factors.

\end{itemize}

The models we discussed thus far are static, that is, they do not
model the change in trust over time. Marsh \cite{Marsh1994a} also
presented a notation for change in trust as an update function for
Equation~\ref{eq:Trust}:
\begin{equation}
    T_x(y,\alpha)^{t+1}=T_x(y,\alpha)^t+ \Delta
    \label{eq:Delta}
\end{equation}

Marsh \cite{Marsh1994a} considered change in trust over time as a
result of different dispositions. For optimists, $\Delta$ is a
large positive number in response to situations with positive
utility, but a small negative number in response to situations
with negative utility. In contrast, for pessimists $\Delta$ is a
small positive number in response to situations with a positive
utility, but a large negative number in response to situations
with negative utility. Different types of optimists and pessimist
can be created by varying the choices of $\Delta$. Jonker and
Treur \cite{Jonker1999} expand on the two trust dispositions by
distinguishing six types of trust dynamics: blindly positive;
blindly negative; slow positive + fast negative; balanced slow;
balanced fast; and slow negative + fast positive. Other models
also assume that trust may change over time due to progressive
accumulation of experiences. In this case,
Equation~\ref{eq:BATrust} becomes:
\begin{equation}
    T_x(x,y)^t=E(B_{x,y}|O_{x,y}^{1:t})
    \label{eq:ProgressiveDelta}
\end{equation}

Change in trust over time, is generally assumed to be a result of
information gathering `experiences'. A parallel concept that is
often discussed as a result of such experiences is confidence
\cite{Griffiths2003,LukeTeacy2006,Griffiths2006}. The intuition is
that the more evidence used to compute trust should result in
higher confidence in computed trust values. Confidence is thus not
a direct measure of trust, but an associated concept that tells an
agent something about its trust calculation.

An example of a confidence metric that can be derived from a
Bayesian definition of trust defines confidence $\gamma_{x,y}$ as
the posterior probability that the actual value of $B_{x,y}$ lies
within an acceptable margin of error $\epsilon$ about $T_{x,y}$.
That is:
\begin{equation}
\Gamma_{x,y} = \frac
{\int\limits_{T_{x,y}-\epsilon}^{T_{x,y}+\epsilon}
w^{\alpha-1}(1-w)^{\beta-1}\mathrm{d}w} {\int\limits_0^1
u^{\alpha-1}(1-u)^{\beta-1}\mathrm{d}u} \label{eq:Confidence}
\end{equation}

An alternative, simpler definition of confidence used in a fuzzy
setting is a simple sum of the number of experiences (positive and
negative) encountered by the agent \cite{Griffiths2006}.

Reliability is an associated concept with trust and confidence.
Reliability theory is the study of the performance of a system of
failure-prone elements \cite{Mahoney2005}. Reliability is studied
in many contexts, including software reliability, network
reliability, hardware reliability and so on. Reliability is
concerned with the ability of computer systems to carry out a
desired operation. Reliability has been understood in different
ways by different researchers with respect to trust. Some
researchers have equated reliability with trust
\cite{Mahoney2005}, while others make the weaker assumption that
reliability influences trust \cite{Fan2008}.

In summary, different trust models have been proposed over the
last two decades, with new models emerging in line with new
developments in machine learning theory. The next section examines
the agent architectures in which these models have been used.

 \subsubsection{Computations by Trust}

Agents are entities that sense their environments using sensors,
reason about data so obtained using one or more reasoning
processes, then act to change their environment using effectors.
While some theoretical work on trust models trust in the absence
of a specific agent architecture, there are some common agent
frameworks that have been used to examine the models. Much of the
literature on trusted agents makes the assumptions that agents are
part of an open multi-agent system (MAS), in which individual
agents are self-interested. That is, each agent has a goal or
goals, but agents in the MAS do not necessarily share a common
goal. This is considered a key reason for the necessity of trust.
Aside from the assumptions of goals and self-interest, trust
models have been used in a range of agent architectures, including
belief-desire-intention (BDI) agents, reinforcement learning
agents and motivated agents. These architectures, and the way in
which they incorporate aspects of trust, are reviewed as follows:

\begin{itemize}

\item \textbf{Belief-Desire-Intention Agents:} The BDI model
\cite{Wooldridge2000} enables the definition of intelligent agents
that have:

\begin{itemize}
    \item Beliefs about the world,
    \item Desires that they would like to achieve,
    \item Intentions, subsets of desires to which they have made some commitment,
    \item Plans that, if successful, will achieve intentions.
\end{itemize}

The BDI architectures is one of the longest-standing models of
intelligent agency used in MASs. It has thus been a popular target
for the exploration of trust models
\cite{Jarvis2006,Taibi2010,Koster2013}. One example is the
ability-belief-commitment-desire (ABCD) model of trust
\cite{Jarvis2006}, which integrates with the logic of rational
agents (LORA) BDI agent theory. Koster et al. \cite{Koster2013}
provide a methodology for incorporating multiple different trust
models BDI agents.

\item \textbf{Motivated Agents:} Griffiths and Luck
\cite{Griffiths2003} extend a BDI framework with motivations. In
addition to the traditional components of the BDI model however,
Griffiths and Luck \cite{Griffiths2003}  argue that motivation is
an extra component required to achieve true autonomy in such
agents. Motivations are high-level desires that characterise an
agent; they guide behaviour and, at a fundamental level, control
reasoning. They argue that short-term teams are not appropriate
since motivations are too dynamic when considered over a short
period of time, i.e. over a small number of tasks; in the
short-term agents' motivations may be out of step resulting in
agents' failure to share common goals at the same time. Similarly,
long-term coalitions are unsuitable since although a given agent
has a fixed set of motivations, the general trend of which
motivations are active may change.

\item \textbf{Learning Agents:} Learning agents can modify their
internal structures in such a way to improve their performance
with respect to some task. Different approaches to learning
existing, including learning from examples (supervised learning)
and learning from reward and punishment in response to
interactions with an environment (reinforcement learning). Many
agents that incorporate models of trust are learning agents,
because they accumulate experiences (examples) and change their
interaction partners over time to improve their performance at
some task. However, some approaches also use secondary learning
algorithms to augment the trust model. Aref and Tran
\cite{Aref2014}, for example, describe a fuzzy trust model in a
Q-learning reinforcement learning framework to help trust
evaluating agents select beneficial trustees for interaction in
uncertain, open, dynamic, and untrusted MASs. Experimental results
indicate that the proper augmentation of a fuzzy trust subsystem
to Q-learning can be useful for trust evaluating agents, and the
resulting model can respond to dynamic changes in the environment.

\item \textbf{Multi-Agent Optimisation:} Trust has also been
studied in optimisation settings such as particle swarm
optimisation (PSO). In PSO, agents (particles) represent solutions
to a problem. Agents move around in a search space attempting to
discover and converge on optimal solutions to the problem in
question. Xu et al. \cite{Xu2012} proposed a variation of the PSO
algorithm for trust path selection in networks. They do not define
a model of trust, but rather a novel PSO algorithm that can embed
a given model of trust.  Huang et al. \cite{Huang2013} propose a
different variation of PSO for incorporating trust to solve grid
task scheduling. PSO algorithms are also particularly compatible
with the trust-region formalism, and examples of algorithms that
incorporate trust-regions with PSO have been proposed
\cite{Merrick2009}.

\end{itemize}

The examples above demonstrate that trust models have been
developed across a number of different agent frameworks. These
include learning, optimisation and planning architectures. Trust
models have been particularly well studied in BDI agents.
Currently we see that the trend is to develop different kinds of
trust models for use in different settings. That is, there is
currently no universally accepted trust model appropriate for a
range of different types of agent frameworks. However, there are
some trust models that have been used and adapted in several
different settings. Marsh's early work \cite{Marsh1994a} is one
such example. These models are general enough for adaptation to
specific agent frameworks or application domains.

Applications of trust models include a range of abstract scenarios
modelled in simulation, as well as a number of well known,
real-world applications. Abstract applications include delegation,
cooperation and detection of deceitful agents. Real world
applications include a range of internet mediated service
providers. A number of applications are discussed below

\begin{itemize}

\item \textbf{Delegation:} Trust models can be used to determine
whether a task should be delegated to another agent. That is,
trust is used by agent $x$ to decide if it should allocate a
certain task to agent $y$ \cite{Burnett2011}.

\item \textbf{Cooperation in Multi-Agent Systems:} Trust can also
be used to determine whether multiple agents should cooperate.
Griffiths and Luck \cite{Griffiths2003} describe different types
of cooperative groups that can occur between agents:

\begin{itemize}

\item Teams: short-term cooperative groups formed to achieve a
specific goal. Agents hold a common goal at the time of forming
the team, and expect immediate benefit from achieving the goal.

\item Clans: medium-term cooperative groups that enhance
individual goal achievement, with increased group performance a
useful side effect.

\item Coalitions: long-term cooperative groups formed to achieve a
specific goal, which can be broken into sub-goals. Agents may have
different preferences for these sub-goals, but are willing to
forgo immediate individual benefit in the interests of long-term
benefit for the coalition.
\end{itemize}

Marsh \cite{Marsh1994} outlines different ways in which trust
plays a part in the formation of cooperative groups:

\begin{itemize}
    \item Trust may influence whether an individual desires to join a group.
    \item Trust may influence whether a group desires an individual to join them.
    \item Trust may influence how members of different groups interact with members of other groups.
\end{itemize}

\item \textbf{Detecting Deceitful Agents:} Deceit is considered
commonplace in certain industries such as trade
\cite{Jonker2005,Sun2005,Hofstede2006,Hofstede2009}, computer
networks \cite{Das2011} and online communities \cite{Xiong2004}.
Understanding deceit is vital for detecting it, and for the design
of governance mechanisms to discourage deceit. Observed
regularities related to trust can be used to form the basis of
models for deceit. This can assist with the detection of deceit in
trade scenarios
\cite{Jonker2005,Sun2005,Hofstede2006,Hofstede2009}, or malicious
behaviour on computer networks \cite{Xiong2004,Das2011}.

\item \textbf{Reputation:} Reputation models are increasingly
ubiquitous among internet mediated service providers
\cite{Josang2007}. For example, eBay (www.eBay.com) uses a
reputation model to rate the trustworthiness of sellers. AirBnB
(www.airbnb.com.au), use reputation models to rate the
trustworthiness of hosts.

\end{itemize}

An appropriate use of trust models necessitates mechanisms for
evaluating the performance of these models. Yu et al.
\cite{Yu2013} include in their survey of trust models a survey of
performance assessment methods for such models. They note two
major approaches used in the literature:

\begin{enumerate}
    \item Simulation-based evaluation; and
    \item Evaluation through real world datasets.
\end{enumerate}

The approaches have been applied either individually or in
combinations by researchers. They found that the most widely used
evaluation method is via simulations. An example of a simulation
testbed is the agent reputation and trust (ART) testbed
\cite{Fullam2005}, which simulates painting appraisers with
varying levels of expertise in different artistic eras. Clients
request appraisals for paintings from different eras. Appraising
agents may also request opinions from other appraising agents.
Appraisers receive more clients, and thus higher profits, for
producing more accurate appraisals. The testbed assesses the
quality of appraisers' trust-based decisions objectively by the
profits they achieve, as well as the accuracy and consistency of
their appraisals.

Simulations permit researchers to vary experimental conditions.
For example, they may simulate different ways in which a trust
model may be exploited. In contrast, real world data may enable
researchers to have a better idea of how their models would work
under realistic environment conditions. However, because many
datasets are not specifically collected for the purpose of
evaluating trust models, they may lack the ground truth about the
user behaviour to facilitate more in-depth analysis of the
performance of proposed trust models \cite{Yu2013}.

\begin{figure*}[!ht]
 \centering
 \captionsetup{justification=centering}
 \includegraphics[width=12cm,height=8cm]{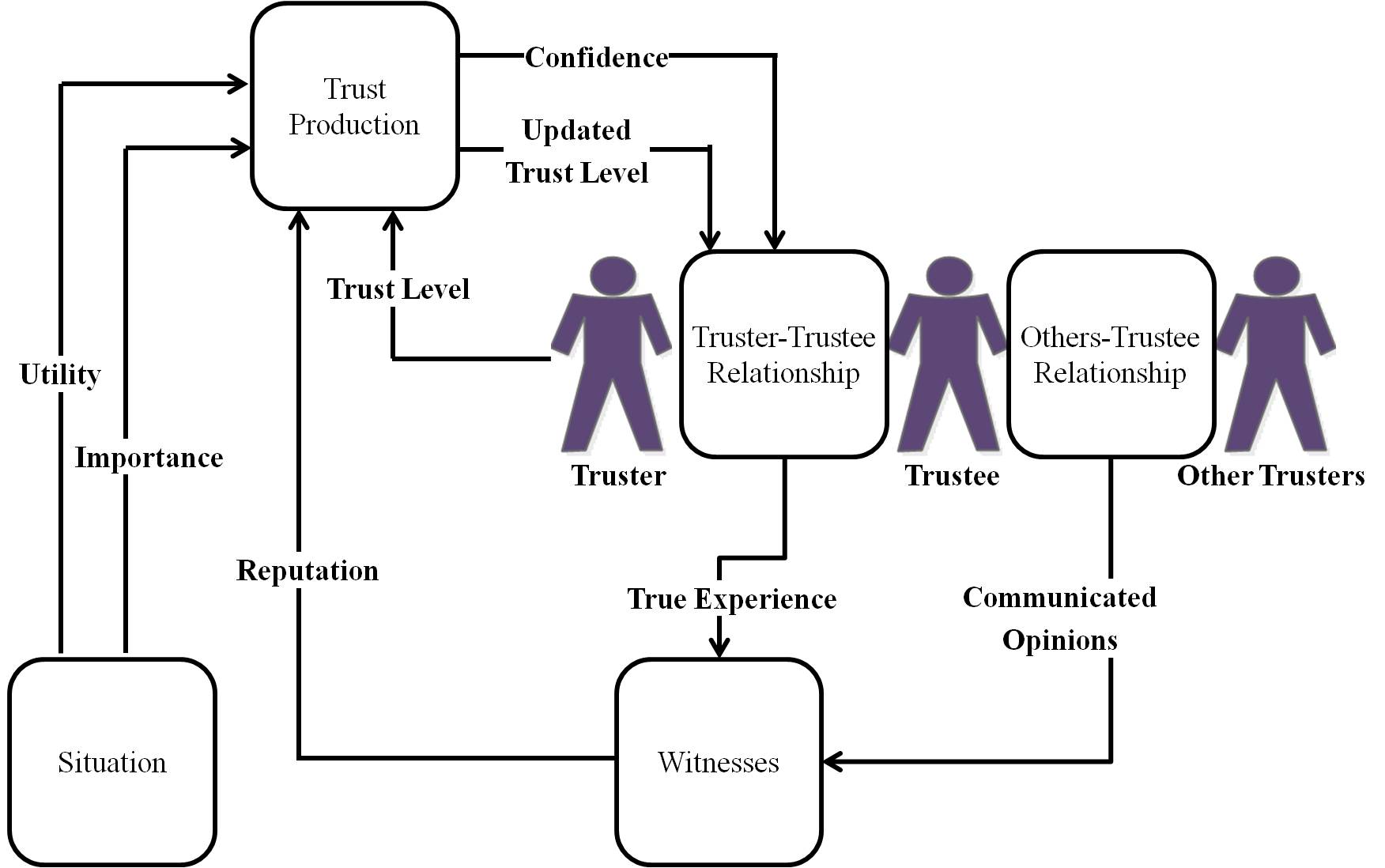}
        \caption{A schematic diagram of different components of a trust model based on the literature reviewed above.}
        \label{fig:AbstractIO}
\end{figure*}

 \subsection{Synthesis of Computational Trust}

This section has considered four aspects of trust models:

\begin{itemize}
    \item The inputs, processing types and outputs of trust models,
    \item The agent architectures in which trust models have been used,
    \item Applications of computational trust,
    \item Techniques for the evaluation of trust models.
\end{itemize}

Regarding the inputs, processing types and outputs of trust
models, a key division is the direct and indirect input types.
Indirect input is frequently indicative of a reputation based
model of trust. Trust has been incorporated extensively in BDI
agent architectures, but has also been examined in motivated
agents and learning agents. Abstract applications include
delegation, cooperation and detecting deceit in MASs. Concrete
applications occur particularly in internet mediated service
industries, which tend to make use of reputation models and
computer networks for security purposes.

Figure~\ref{fig:AbstractIO} presents a compilation of the
literature on trust modelling that was reviewed in the previous
section.

This diagram clearly identifies the four main inputs to a trust
model. One input depends on the trustor's initial belief about the
level of trust it holds towards the trustee. One input depends on
direct and indirect experience encounters with the trustee. Two
inputs depend on context and represent the importance and value of
the context.

The trustor needs to obtain two indicators from the trust model:
the updated level of trust and the model confidence in this
update. The trustor may choose to ignore an estimate with low
confidence.

This model is intuitive, simple and basically captures the basic
building blocks from across the wide literature on modelling
trust. Variations can exist by expanding some of the elements. For
example, how to calculate the importance of a situation or what
personal attributes of the trustor and trustee we can use for
trust update. Another example is related to the risk appetite and
attitude of the trustor, which may influence the trustor's
trust-update functions. All these are possible, with their utility
dependant on data availability and the cost-benefit trade-off that
needs to be taken into account when deciding on how complex a
trust-model should be.

Trust production is the set of mechanisms needed to make a
trusting decision. Figure~\ref{fig:AbstractIO} conceptually
depicted one input-and-output view of a trust production system.
It is equally important to know where trust sits within the wider
agent architecture. We will start with the BDI architecture since
we reviewed it above, but we will then move high up in the
abstraction hierarchy to offer system-level architectures that
integrate trust production more profoundly than the simple
equation model used in almost all of the literature we reviewed on
trust.

\begin{figure*}[!t]
 \centering
        \captionsetup{justification=centering}
        \includegraphics[width=15cm,height=10cm]{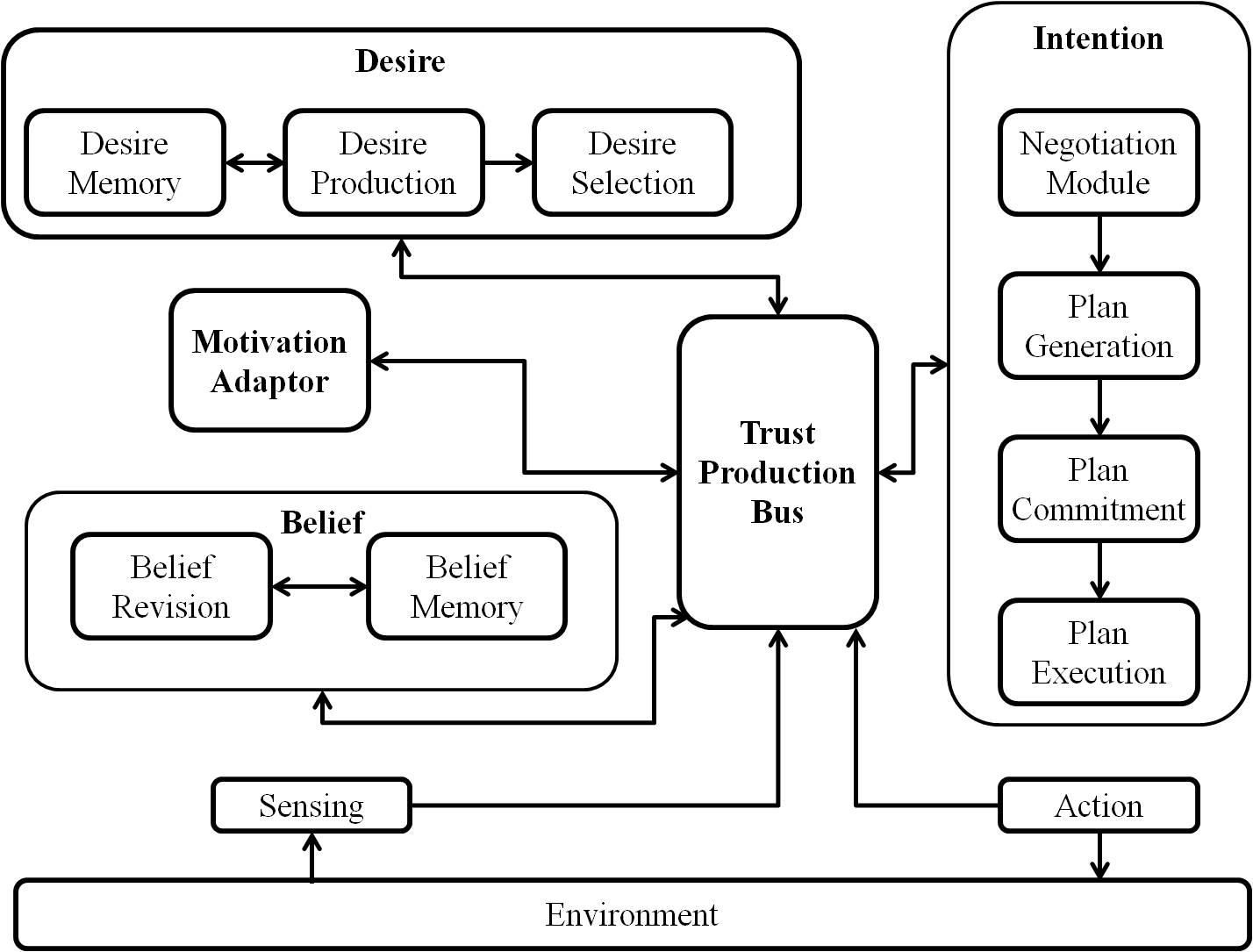}
        \caption{A schematic diagram of a modified BDI agent.}
        \label{fig:AbstractAgents}

\end{figure*}

The amalgamation of trust production with BDI necessitates a new
perspective on BDI that explicitly acknowledges the complexity of
trust production instead of merely reducing it to the
equation-based approach. We compile the above discussions into the
Belief-Desire-Intention-Trust-Motivation (BDI-TM) architecture as
shown in Figure~\ref{fig:AbstractAgents}.

This architecture acknowledges two additional components to BDI;
these are: Trust Bus and motivation adaptor. The Trust Bus
interacts with the belief component to estimate trustworthiness of
trustees but also produces information which are essential for the
belief update mechanism: the primary learning mechanism in a BDI
architecture. The Trust Bus acts as the message passing interface
that learns -- through automatic knowledge acquisition
methods~\cite{Leu2016} -- from and modifies massages as they get
transferred from one system module to another.

The second component is related to motivation, which acts as an
adaptor to desires. We prefer in this architecture to make
motivation as a stand-alone component, simply because it is an
adaptor to the overall desire production box, not just the desires
themselves and can influence, and be influenced with, trust
production as well. Beliefs memory and trust need to get fused
with motives before motives adapts desires.

\begin{figure*}[!t]
 \centering
 \captionsetup{justification=centering}
 \includegraphics[width=12cm,height=8cm]{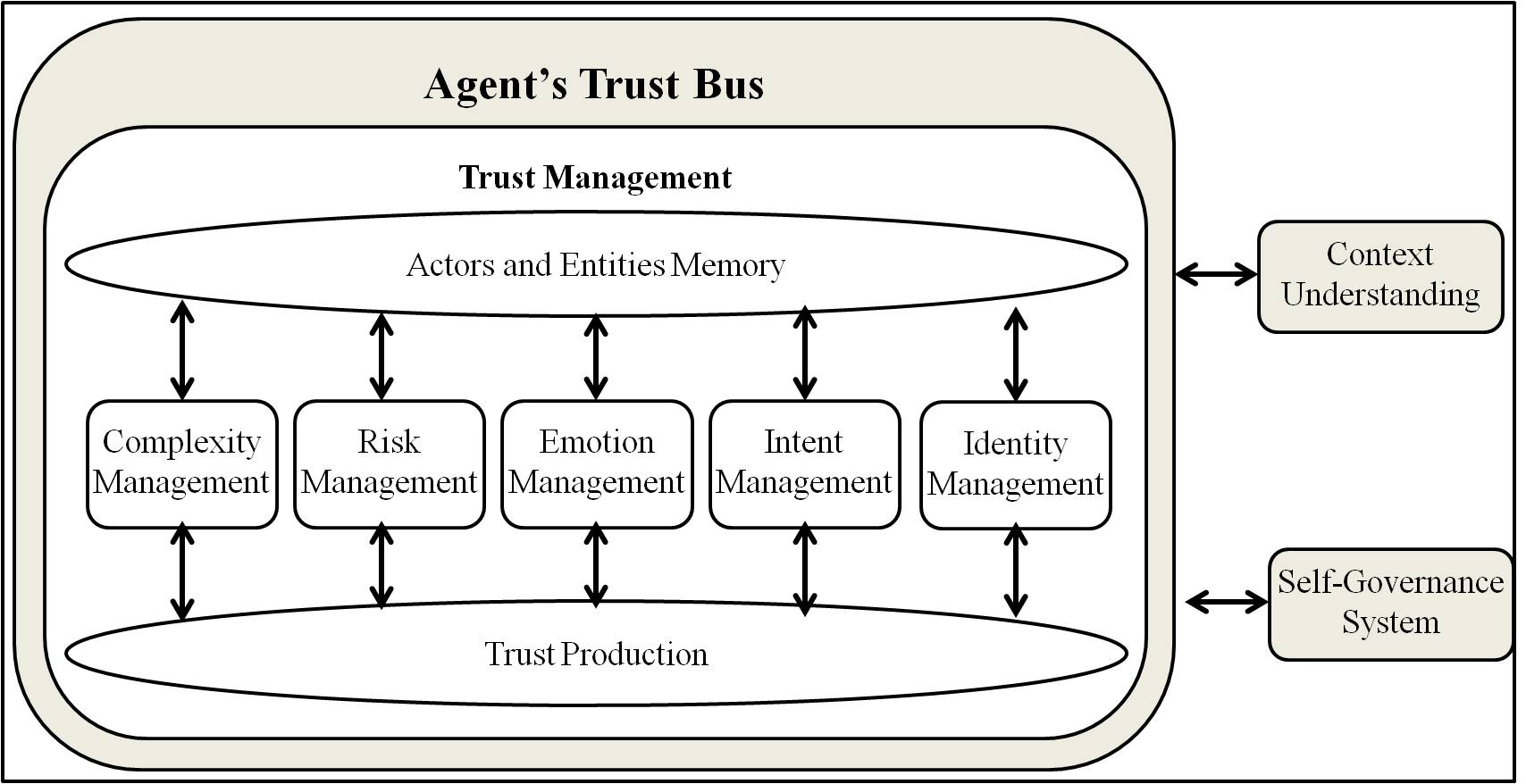}
 \caption{TrustBus.} \label{TrustBus}
 \end{figure*}

The previous discussion begs question on the constituents of the
Trust Bus, and whether this bus is a simple message-passing
interface or does it do more than that?

Figure~\ref{TrustBus} expands the Trust Bus into its constituents:
Actors and Entities Memory, Trust Production, Identity Management,
Intent Management, Emotion Management, Complexity Management and
Risk Management. Each of these components deserves a brief
overview below.

The literature on trust implicitly emphasized the need to be aware
of the actors in the environment (at least the trustee) and other
entities. The Actors and Entities memory maintains the information
stored on these actors and entities for retrieval when needed. The
identity management module is responsible for identification of
actors and continuous authentication of their identities while
interacting with the system. This is a critical component
underpinning the mechanisms needed to manage reputation,
reliability of information, privacy and security.

The intent management module acts as a smart engine to objectively
estimates actors intents and use these estimates to offer informed
advice to the trust production module on the reliability and value
of these intents. It also offers what-if services to the trust
production module to explore the space of possible implications of
trust.

The three remaining modules represent the cornerstones for trust
in psychology and social science research and were elicited in our
previous
work~\cite{petraki2014trust,abbass2015n,abbass2015trusted}. These
are emotion, risk, and complexity. Emotion management models and
learns affective states of other actors in the environment. An
advanced version of this module can manage the emotion states
expressed by the AI when needed. The risk management module
maintains a risk registry, analyzes uncertainties and offer
what-if services to the trust production module. The complexity
management module estimates the cognitive and task complexities in
the context of other agents and the trustor. The literature on
trust shows that complexity is a critical driver for trusting
decisions. The trustor can get exposed to a level of complexity
forcing unwise trusting decisions if this complexity is taken for
grant. The complexity management module analyses the complexity
surrounding the agent in a given situation, including the
complexity surrounding other actors in the environment and their
abilities to manage it.

The Trust Bus is used to scope the discussion on the nature and
types of sensor technologies required by the Trust Bus to estimate
human and group states. The Big Data nature of this problem
warrants a special treatment and raises both theoretical and
practical challenges. This topic will be the context of the next
section.

 \section{Big Data at the Human-Machine Interface}\label{BDHMI}

The machine needs to be able to sense humans, their behavioral
attributes, cognitive states, actions and decisions, and even the
rational and explanations they make about their own judgements.
This sensing is key for the machine to acquire the necessary
information it needs to make a judgement about a human. It is very
similar to human-to-human interaction, where humans actively and
subconsciously sense each others facial expressions, body
language, voice tone, and even skin temperature during a hand
shake or a hug. With today's technologies, such as wearable
devices, the machine can access even more information about the
human's psychophysiological states that the humans themselves may
not be conscious of. Today, Google applications can map out
humans' daily routines, food and cloth preferences based on their
shopping style, estimating probability of heart attacks based on
their life style, social networks based on people in their social
circuit, similar locations at night, similar routes, and more.

In addition to sensing the human, the machine needs to sense the
environment. A change in human body temperature can simply be
attributed to changes in the air conditioning setting. A change in
a human voice signal can be attributed to the temperature in the
environment, interferences between something in the environment
and the microphone, or simply a result of a long speech or
fatigue. Without sensing the environment, the machine is unable to
attribute changes in the signal that it is sensing to different
possible human states.

Sensing is a pre-requisite for any modelling that a machine needs
to estimate a human's trust level. This section covers this topic
through a Big Data lens. Before we regress further, we start the
discussion with how big is a big data problem.

 \subsection{How Big is Big}

How big is big? This is usually the question that gets asked by
those who are skeptical about the word `big' in big data. However,
the word `big' is not merely about volume. It reflects the size of
different characteristics that together form the basic features
that one can use to characterize a problem as a big data problem.

The common characteristics for big data are summarized in
different words that start with the letter `V'. This generated an
evolution of these characteristics starting with the 3Vs model
representing volume, velocity, and variety, then the 5Vs with the
addition of veracity and value, and finally the 6Vs with the
addition of
variability~\cite{russom2011big,jin2014computational,abbasscrt1}.

Each of the Vs represent one characteristics that by itself, the
problem may get called a big data problem, but the challenge gets
more complex when multiple Vs coexist together. Naturally the
first V represents the volume or size of the data. A big data
problem may have a significantly large number of features or
simply terabytes or petabytes files.

The second V stands for velocity, or the speed with which the data
change. Such changes may cause movements of the decision
boundaries for a decision tree, alterations of the degree of
nonlinearity in a regression model, or variation in the prototypes
and numbers of clusters in a clustering problem.

Variety, or heterogeneity, refers to the distributed nature of the
data. A customer may have different forms of records across
different databases. Some of the records may take the form of
buying patterns, while others may take the form of voice
signatures to authenticate the customer during online banking, or
unstructured data collected during an interview with the customer.

Veracity is related to the amount of noise that may exist in the
data and the level of trustworthiness of the source. Veracity can
further be divided into accidental veracity and intentional
veracity. The former represents the classic noise in a
communication channel associated with natural and/or physical
causes. The latter represents intentional noise caused by a
malicious act. These two factors play an important role in the
validity and authenticity of any conclusion drawn from a dataset.

The second last V is the value characteristic; capturing the value
for money and worthiness of the decision or conclusion drawn from
the data. This is a cost-benefit-analysis question. Low cost data
can generate high benefits, while high cost data may be too costly
for the benefits they generate.

Last, but not least, variability represents the dynamic nature of
most IT infrastructure. Data exchange protocols change. The
architecture of a database changes. The data format received from
third parties change. These changes are necessary as we optimize
the databases, customize the data differently for different
customers, and so forth. Nevertheless, these changes impose major
challenges in big data. The more variability that exists in the
problem at hand, the harder the big data problem is.

Autonomous systems are at the heart of the Big Data problem. The
problem of autonomy is that it can generate unpredictability in
the environment. As a software robot moves from one computer to
another, protocols may change, data format may change, data
worthiness may change, source reliability may change, and so
forth. An autonomous entity has limited processing power relative
to the data it encounters during its operations. Autonomy is a
challenge for big data, and equally though, big data is a
challenge in autonomy.

Adding trust to the equation creates the biggest challenge
overall. Trust glues autonomous systems. This glue is not a
one-off glue. Trust is dynamic and requires dynamic strategies to
manage it in an autonomous system.

 \subsection{Sensors: What can they offer the human machine problem}

Sensing is what allows a system to interact with its environment
in a flexible and non-deterministic way, as apposed to performing
preprogrammed actions based on a set of scheduled functions and/or
rule sets \cite{Moron2015}. In other words, sensing is a
fundamental mechanism and condition for a system to be able to
operate autonomously in order to fulfil its purpose. Recent years
have marked unprecedented advances in sensor technology, making
available miniaturised, highly portable, accurate and reliable
sensors, with high throughput and low cost \cite{Bisdikian2007}.
This allowed the resultant sensor-based autonomous systems to
embed more sensors, which in turn allowed them to operate in more
complex and more dynamic environments. On the one hand, one can
consider that the better the sensing is, the more data the system
gathers from the environment, and further, the more informed and
accurate the resultant decision is. However, on the other hand, as
the amount of data generated by sensors grows, so do (1) the
amount of processing the system needs to perform in order to
generate a useful decision, and (2) the difficulty of taking the
decision given the increased amount of choice. Rich data coming
from numerous and high quality sensors could be simultaneously an
enabling and a disabling factor for the operation of autonomous
systems, with respect to the processing power available to the
system.

We can give the example of an autonomous unmanned vehicle, which
must deal with data coming from the activity of human agents
served by the vehicle (e.g. passengers of an autonomous car, or
passengers of an aircraft), and also from the surrounding
technological agents (internal interaction with own mechanical
sub-systems or external interaction with other vehicles
participant in traffic). Another example can be an autonomous
system for patient diagnosis and monitoring in a hospital. This
needs real-time and historical data from the respective patient
and historical data from numerous other patients in order to
ensure statistical validity of the decision. Also, it needs data
from the various machines and equipment operating in the hospital.

Regardless of the source and type of activity, modern sensors
operate in a digital manner, and generate data through sampling of
their sensing. While sensing and processing of sensory information
are taken for grant for biological agents (e.g. a human
observer/decision-maker), it is not the case with artificial
autonomous systems, which operate through digital computation and
perceive everything as data. Thus, it becomes essential to have a
clear view of the existing sensor technologies especially with a
focus on the amount of data they generate as part of their normal
operation. At the same time, it is essential to have a clear view
on the existing processing/computing technologies and
architectures in order to understand the potential limitations the
autonomous systems can experience when they deal with the big data
coming from their sensory-subsystem.

\subsubsection{Sensing human activity}

Human activity sensors have undergone unprecedented development in
recent years. Sensors and sensing systems that were once reserved
for clinical or research applications are now easily accessible to
the general public; are manufactured and commercialised on a large
scale; and are being used across multiple domains in versatile and
low cost applications. Patel et al. \cite{Patel2012} identified
the key facilitators that enabled this explosion of achievements
in sensing human activity. The authors discuss the sensing systems
in terms of three essential building blocks: (1) the sensors, i.e.
sensing and data collection system (2) the communication system
needed to transfer data to the processing centre, and (3) the data
analysis techniques that extract the relevant information from the
raw data. A refinement of the three component view presented in
\cite{Patel2012} was later discussed by Mukhopadhyay
\cite{Mukhopadhyay2015}, who considers eight components where
sensing research should concentrate for improving the overall
performance of human activity wearable sensing systems: (1) types
of sensors to be used, (2) type of communication protocols to be
employed, (3) type of activities to be monitored, (4) techniques
for extracting important features from activities, (5) design and
development of small, light-weight, powerful and low-cost smart
sensor nodes, (6) harvesting of energy for normal operation and
communication, (7) ability to be used with the present day mobile
devices, (8) potential to be reconfigured for new
applications/users.

Both studies, \cite{Patel2012,Mukhopadhyay2015}, note how each of
these components have been on a path of miniaturisation, where the
advances in micro/nano-electronics and related technologies
allowed the resultant systems to become smaller and more powerful,
and in the same time more energy efficient. These technological
advances further allowed the emergence of fully wearable sensing
systems that provide versatility and flexibility in both design
and operation; in addition, they allow long-term non-intrusive
monitoring and data collection. In \cite{Mukhopadhyay2015} the
author states that this surge of wearable sensing will continue to
expand towards more and more domains, and cites industry reports
concluding that the penetration of research advances towards the
consumer market will lead to an estimated market growth to a value
of over \$20 billion by 2017. Wearable sensing systems have been
extensively used in domains such as sports and recreation
\cite{Athos2016}, human-computer interaction, work-space and
ergonomics, games and entertainment, and last but not least,
education.

Numerous reviews discussed sensors and sensing systems in all
these domains from the points of view of their component blocks
and subsequent technology, with a focus on the aspects that
allowed miniaturisation and wearability. Extensive and very recent
reviews can be found in \cite{Patel2012}, \cite{Banaee2013},
\cite{Mirzaei2014}, \cite{Bandodkar2014}, \cite{Stoppa2014},
\cite{Mukhopadhyay2015} and \cite{Sprager2015}. However, the
resultant amount of data was not discussed so far in the
literature. In the following, we describe the most common types of
sensing systems used in monitoring human activity
\cite{Mukhopadhyay2015}, and we focus our discussion on the
sampling rate and the subsequent data resulting from their
operation.

\begin{itemize}

\item \textbf{Body temperature:} Body temperature is one of the
most common physiological parameters used for monitoring human
activity, usually measured by wearable sensors attached to the
skin. The body temperature in different parts of the body, and
variations of it can give an indication of symptoms of medical
stress that signal the existence, or the possibility of acquiring
various health conditions. Body temperature monitoring can be used
either for determining the respective physiological/medical
conditions or for classifying and storing for further use
historical physiological activity data, or even for harvesting
energy from body heat \cite{Mukhopadhyay2015}. Traditional
measurement of body temperature was in the form of one-off
measurements, for clinical purposes. However, continuous
monitoring is nowadays common, especially as part of wearable
activity monitoring devices and systems. In
Table~\ref{table:Movement} all devices listed as incorporating
temperature sensors, refer to skin temperature, and operate
through contact to skin tissues typically in the wrist area, where
wristband or watch type activity monitors are attached. Contact
technologies such as thermocouples, resistive temperature
detection, and organic diodes have been proposed over time
\cite{Jeon2013}. Recently, contactless temperature sensors based
on infra-red waves have been developed for accurate and hygienic
monitoring of temperature in various parts of the body, such as
FT90 contactless clinical thermometer from Beurer
\cite{Beurer2016}.

\item \textbf{Heart rate:} Heart rate (HR) is also one of the most
common physiological parameters monitored using wearable sensors
for various purposes. Heart rate is a precisely regulated variable
that is critical for diagnosing various medical conditions and
disease of human, however its monitoring is also essential in
sports and entertainment. A variety of sensor modalities are
available for heart rate measurement and monitoring, such as sound
based \cite{Kovacs2000,Zhu2003Patent}, optics
(photoplethysmography) based \cite{Alzahrani2015,Salehizadeh2016}
or pressure based \cite{Park2015}; also monitoring by visual cues
such as the face colour \cite{Mukhopadhyay2015} have been also
reported in the literature. Historically, the early wearable HR
monitoring devices were pressure-based and used chest-straps to
attach to upper body, e.g. Garmin HRM-Tri and HRM-Swim
\cite{GarminSwimtri} dedicated to triathlon and swimming athletes,
respectively. Recently, the due to the high demand on the wearable
consumer market, technology migrated towards optic-based devices,
which are currently implemented on all wristband, smartwatch,
pendant or garment activity sensors. In Table~\ref{table:Movement}
all activity sensing devices that include HR monitoring are
LED-enabled optic-based sensors. An in-depth discussion on the
differences between the two technologies, and the trends in
wearable HR devices is provided in \cite{RAF2016}.

\item \textbf{Electroencephalography:} Electroencephalography
(EEG) is an electrophysiological sensing method that captures
brain activity in the form of brainwaves coming from the
aggregation of the electrical neural impulses; thus, EEG is based
on electromagnetic wave sensing. Typical EEG sensing systems
consist of several electrodes that measure in different regions of
the brain the voltage fluctuations resulting from ionic current
within the neurons. The sensing can be invasive, where the
electrodes are implanted under the scalp, or non-invasive, where
the electrodes are placed on the scalp either individually or as
part of customised caps \cite{Geodesic400EGI} in which the
electrodes are precisely positioned. However, with the recent
advances in the transducing elements of EEG devices, intrusive
methods are now less and less used, while wearable caps and the
very recent dry sensing headsets are becoming more common. EEG
sensing was traditionally used in clinical applications, such as
the traditional diagnose method for epilepsy, but also in
supporting the  diagnosis, monitoring, and treatment of other
brain conditions, such as sleep disorders, coma, encephalopathy,
brain death, etc. Research applications are especially in the
fields of cognitive science and engineering \cite{Abbass2014a},
and human/brain-computer interaction \cite{Abbass2014}, where the
high temporal resolution of EEG sensing brings valuable
information about real-time brain activity, capturing changes in
cognitive load and states. High temporal resolution is the main
advantage of EEG over a number of other brain imaging and brain
activity sensing methods, which include magnetic resonance imaging
(MRI) and functional MRI, single-photon and positron emission
tomography, magnetoencephalography (MEG), nuclear magnetic
resonance spectroscopy, electrocorticography, near-infrared
spectroscopy, and event-related optical signaling. Classic
clinical EEG sensing systems operate at sampling rates between 100
and 256 Hz \cite{EmotivInsight,EmotivEpoc,NihonkohdenTrackit},
while some of the latest EEG systems are capable of recording at
sampling rates as high as 20 kHz
\cite{SynampsCompumedics,eegoMylabantNeuro,GraelCompumedics}, that
is, a sub-millisecond temporal resolution. However, while this
temporal resolution is the main advantage of EEG sensing, it is
also one of the challenges for using it in real-time applications,
given that EEG systems can record in parallel on as many as 256
channels. In Table~\ref{table:EEG} we summarise the main EEG
equipment available to date, and list for each of them the
essential sensing data-generator features.

\begin{table*}[t]
    \centering
    \caption{Summary of EEG sensing equipment.}
    \label{table:EEG}
    \begin{tabular}{|c|c|c|c|c|c|c|c|}
        \hline
        \textbf{Producer} & \textbf{Model} & \textbf{Grade} & \textbf{Mobility} & \textbf{Frequency range} & \textbf{Channels} & \textbf{Sampling rate/channel} & Reference \\
         & & & & \textbf{[Hz]} & & \textbf{[Hz]} & \\ \hline

        EGI & Geodesic 400 & clinical & fixed & N/A & 32-256 EEG & 1000 & \cite{Geodesic400EGI} \\ \hline
        Mindmedia & Nexus32 & research & fixed & N/A & 24 EEG & 2048 & \cite{MindMedia2016} \\ \hline
        Mitsar & EEG-202 series & research & fixed & 0-150 & 24-31 EEG & 2000 & \cite{Mitsar2016} \\ \hline
        Nihon Kohden & Neuroxax 1200 & research & fixed & 0-1200 & 256 EEG, 16 DC & 10000 & \cite{NohonkohdenNeurofax} \\ \hline
        Compumedics & GraelHD & research & fixed & 0-3500 & 4 $ \times$ 64 EEG & 20000 & \cite{SynampsCompumedics} \\ \hline
        Neurovirtual & BW series & clinical & fixed & 0-500 & 25-50 EEG, 8 DC & 2000 & \cite{BW3Neurovirtual} \\ \hline

        Nihon Kohden & Trackit & clinical & mobile & 0-100 & 32 EEG, 4 DC & 200 & \cite{NihonkohdenTrackit} \\ \hline
        NeuroConn & Thera Prax & research & mobile & 0-1200 & 22 EEG & 32-4000 & \cite{Neuroconn2016} \\ \hline
        EGI & Geodesic 100 & clinical & mobile & N/A & 32 EEG & 1000 & \cite{Geodesic100EGI} \\ \hline
        Medicom MTD & encephalan EEGR & clinical & mobile & 0.016-70 & 20 EEG & 2000 & \cite{Medicom2016} \\ \hline
        Neurowerk & Neurowerk DB series & clinical & mobile & N/A & 21-25 EEG & 128-512 & \cite{Neurowerk2016} \\ \hline
        Compumedics & GraelHD & clinical & mobile & 0-580 & 32 EEG & 16000 & \cite{GraelCompumedics} \\ \hline
        Compumedics & Nuamps & clinical & mobile & 0-260 & 40 EEG & 1000 & \cite{nuampscompumedics} \\ \hline
        antNeuro & eego mylab & clinical & mobile & N/A & 32-256 EEG & 16000 & \cite{eegoMylabantNeuro} \\ \hline
        Deymed & TruScan series & research & mobile & 0.16-450 & 24-256 EEG & 4000 & \cite{DeymedMobile} \\ \hline

        Neurowerk & Neurowerk EEG mobile & clinical & wearable & N/A & 16 EEG & 128-512 & \cite{Neurowerk2016} \\ \hline
        antNeuro & eego sports & clinical & wearable & N/A & 32-64 EEG & 2000 & \cite{eegoSportsantNeuro} \\ \hline
        Neurovirtual & BWmini & clinical & wearable & 0-500 & 33 EEG, 2 DC & 1000 & \cite{BWminiNeurovirtual} \\ \hline
        NeuroSky & BrainWave & research & wearable & 0-100 & 1 EEG & 512 & \cite{NeuroSky2016} \\ \hline
        Mitsar & EEG-201 series & clinical & wearable & 0.16-70 & 21 EEG & 500 & \cite{Mitsar2016} \\ \hline
        Emotiv & Epoc & research & wearable & 0.2-45 & 14 EEG, 2 ref & 128 & \cite{Louwerse2012,Fraga2013,EmotivEpoc} \\ \hline
        Emotiv & Epoc+ & research & wearable & 0.16-43 & 14 EEG, 2 ref & 256 & \cite{EmotivEpoc} \\ \hline
        Emotiv & Insight & hobby & wearable & N/A & 5 EEG, 2 ref & 128 & \cite{EmotivInsight} \\ \hline
        Macrotellect & BrainLink & hobby & wearable & 8-30 & N/A & N/A & \cite{Macrotellect2016} \\ \hline
        Interaxon & Muse & hobby & wearable & N/A & 7 EEG & N/A & \cite{Charara2015,Interaxon2016} \\ \hline
        Somnomedics & SOMNOsecren EEG 32 & research & wearable & N/A & 25 EEG & 4-512 & \cite{Somnomedics2016} \\ \hline
        Neurosoft & Neuron Spectrum AM & clinical & wearable & N/A & 21 EEG & 1000 & \cite{Neurosoft2016} \\ \hline
        Deymed & TruScan wireless & research & wearable & 0.01-1200 & 32-256 EEG & 6000 & \cite{DeymedMobile} \\ \hline
        Deymed & BFB & clinical & wearable & 0.3-70 & 2 EEG & 256 & \cite{DeymedBFB} \\ \hline
    \end{tabular}
\end{table*}

\item \textbf{Eye movement sensing:} Sensor systems that capture
eye movement are used in clinical contexts primarily in
ophthalmological diagnosis and in research contexts for eye
tracking applications mainly related to human factor, cognitive
science and engineering, and human-computer and brain-computer
applications. The common eye movement sensing systems are either
based on electric sensing or optical sensing.

The most common electric sensing method is Electrooculography,
which measures the corneo-retinal standing potential existing
between the front and the back of the human eye. To measure eye
movement, pairs of electrodes are placed at the opposite sides of
the eyes (i.e. either above and below the eye or to the left and
right of the eye). If the eye moves from centre position toward
one of the two electrodes, this electrode ``sees'' the positive
side of the retina and the opposite electrode ``sees'' the
negative side. The resulting signal is called electrooculogram, in
which the difference in potential occurring between the electrodes
is recorded as a measure of the eye's position. EOG captures eye
movement only, while more complex electric sensing methods, such
as the Electroretinogram, can also capture and monitor eye
response to individual visual stimuli.

The optical eye movement sensing typically consists of
eye-tracking systems based on specialised video cameras that
capture retinal reflection and refraction as result of eye's
exposure to visual stimuli.

\item \textbf{Electrocardiography:} Heart activity can be also
measured and monitored using sensing systems. The typical sensing
method for heart activity monitoring is electrocardiography (ECG),
which uses electrical sensing for detecting the electrical changes
that arise from the heart muscle depolarising during each
heartbeat. These changes are captured on subject's skin using
electrodes placed on relevant locations on the body. Typical ECG
systems, used in clinical contexts, consist of ten electrodes
placed on patient's limbs and chest, which capture over various
periods of time the overall magnitude of the heart's electrical
potential from twelve different polarisation angles, also known as
``leads''. The output of the sensing is the electrocardiogram,
which contains magnitude and direction of the heart's electrical
depolarisation at any moment throughout the cardiac cycle.

In clinical contexts ECG sensing provides important information
about the physical structure of the heart and the operation of its
electrical conduction system (e.g. rate and rhythm of heartbeats,
size and position of heart chambers, operation of implanted
pacemakers etc.) and have been used extensively for diagnosis of
various heart conditions (e.g. abnormal heart's muscle cells or
conduction system, the effects of cardiac drugs). In addition,
recent wearable ECG sensing systems have been extensively used for
permanent monitoring of people with chronic cardiovascular
diseases. For example, in \cite{Mukhopadhyay2015} the author
mentions a compact plaster sensor designed for wearable low cost
cardiac healthcare, which uses a low power high resolution
thoracic impedance variance for ECG monitoring. In
Table~\ref{table:ECGresting} and Table~\ref{table:ECGstress} we
summarise the popular ECG equipment available to date for resting
and stress contexts, respectively, and list for each of them the
essential sensing data-generator features.

\begin{table*}[t]
    \centering
    \caption{Summary of rest testing ECG equipment.}
    \label{table:ECGresting}
    \begin{tabular}{|c|c|c|c|c|c|c|c|c|}
        \hline
        \textbf{Producer} & \textbf{Model} & \textbf{Grade} & \textbf{Mobility} & \textbf{Frequency range} & \textbf{Channels} & \textbf{Sampling rate/channel} & \textbf{Encoding} & \textbf{Reference} \\
        & & & & \textbf{[Hz]} & & \textbf{[Hz]} & \textbf{[bits/sample]} & \\ \hline

        Cardioline & TouchECG & clinical & mobile & 0.05-300 & 12 & 1000 & & \cite{CardiolineAR600} \\ \hline
        Cardioline & AR600 series & clinical & mobile & 0.05-150 & 12 & 500-1000 & & \cite{CardiolineTouchECG} \\ \hline
        TSE & UCARD200 & clinical & mobile & 0.05-150 & 12 & 1600 & & \cite{TSEUCARD200} \\ \hline
        Nasan & Simul-G & clinical & mobile & 0.05-150 & 12 & N/A & & \cite{NasanSimulG} \\ \hline
        Longfian & ECG-1112 & clinical & mobile & 0.05-160 & 12 & 1000 & & \cite{Longfian2016} \\ \hline
        Nasan & ESY-G1 & clinical & mobile & 0.05-150 & 1 (12 serial) & N/A & & \cite{NasanESYG1} \\ \hline
        Nasan & ASAAN 1003 & clinical & mobile & 0.05-150 & 3 (12 serial) & N/A & & \cite{Nasan103} \\ \hline
        Nasiff & CardioCard & clinical & mobile & 0.05-150 & 12 & 250-1000 & & \cite{NasiffResting} \\ \hline
        SpaceLab & CardioDirect 12 & clinical & mobile & 0-15 & 12 & 2000 & & \cite{SpaceLabCardioDirect} \\ \hline
        CustoMed & CustoCardio 100 series & clinical & mobile & 0-500 & 12 & 1000-4000 & & \cite{Customed1xx} \\ \hline
        Kalamed & KES121 series & clinical & mobile & 0.05-165 & 12 & 2000 & & \cite{KalamedKES121} \\ \hline
        Kalamed & KES1000 & hobby & wearable & 0.05-165 & 12 & 2000 & & \cite{KalamedKES1000} \\ \hline
        Labtech & EC-12RM & hobby & wearable & 0.05-150 & 12 & 512 & & \cite{Labtech2016} \\ \hline
    \end{tabular}
\end{table*}

\begin{table*}[t]
    \centering
    \caption{Summary of stress testing ECG equipment.}
    \label{table:ECGstress}
    \begin{tabular}{|c|c|c|c|c|c|c|c|c|}
        \hline
        \textbf{Producer} & \textbf{Model} & \textbf{Grade} & \textbf{Mobility} & \textbf{Frequency range} & \textbf{Channels} & \textbf{Sampling rate/channel} & \textbf{Encoding} & \textbf{Reference} \\
        & & & & \textbf{[Hz]} & & \textbf{[Hz]} & \textbf{[bits/sample]} & \\ \hline

        CustoMed & CustoCardio 200 series & clinical & mobile & 0-500 & 12 & 1000-4000 & & \cite{Customed2xx} \\ \hline
        Kalamed & KES121 series & clinical & mobile & 0.05-165 & 12 & 2000 & & \cite{KalamedKES121} \\ \hline
        Eccosur & ECG-View & clinical & mobile & 0.05-100 & 12 & 256-600 & & \cite{Eccosur2016} \\ \hline
        Thor & Thoriax PC & clinical & mobile & N/A & 12 & 2000 & 12 & \cite{ThorPC} \\ \hline

        SpaceLab & CardioDirect 12S & clinical & wearable & 0-150 & 12 & 2000 & & \cite{SpaceLab12S} \\ \hline
        Cardioline & CubeStress HD+ & clinical & wearable & 0.05-300 & 6+6 or 12, AVG & 1000 & & \cite{CardiolineCubeStress} \\ \hline
        Cardionics & CarTouch & clinical & wearable & N/A & 12 & 1000 & 16 & \cite{Cardionics2016} \\ \hline
        Labtech & EC12RS & clinical & wearable & 0.05-150 & 12 & 1000 & & \cite{LabtechStress} \\ \hline
        Amedtec & CardioPart 12 Blue & clinical & wearable & 0-150 & 12 & 8000 & & \cite{Amedtec2016} \\ \hline
        Thor & Thoriax Home & hobby & wearable & 0.5-40 & 1 (differential) & N/A & 12 & \cite{ThorHome} \\ \hline
    \end{tabular}
\end{table*}

\item \textbf{Skin Conductivity:} Skin conductivity (SC) sensing
is based on electrodermal activity (EDA), in which electric
properties of the skin varies with environmental conditions and
activity performed \cite{Boucsein2012}. Human skin, especially
that from the extremities of the limbs, e.g. fingers, palms, or
soles of feet, display bio-electrical phenomena that are relevant
to diagnosing a variety of psycho-physiological conditions
\cite{Picard2016}. These phenomena can be detected with sensors
that capture changes in the electrical activity between two points
on the skin over time \cite{Thoughttechnology2016}. Traditionally,
electrodermal activity was associated with the galvanic response
of skin tissues (GSR - Galvanic Skin Response), in which skin
resistance (impedance) varies with the state of sweat glands, when
the glands are excited by the activity of sympathetic nervous
system. Thus, galvanic skin response sensors were used to
determine various aspects related to psychological or
physiological arousal. However, recent insights showed that other
sub-dermal tissues and physiological activity can also influence
the electrical properties of the skin, in addition to the
psychologically-induced gland activity. As a result, the concept
of skin conductivity sensing has been enlarged to encompass the
electric potential response (GSP - Galvanic Skin Potential), where
sensors capture the voltage or changes in voltage measured between
two electrodes conveniently located in two different locations on
the skin.

Modern EDA sensing is largely used for capturing complex emotional
and cognitive states \cite{Mendes2009,Picard2016}, in contexts
like anxiety monitoring and control, stress detection, monitoring
epileptic patients with drug resistance, or as part of polygraph
equipment for lie detection in conjunction with other biofeedback
sensors such as heart rate and blood pressure
\cite{EmpaticaEmbrace}, or respiratory rate.
Table~\ref{table:Skin} provides a summary of the most popular skin
conductivity monitoring equipment available to date, and list for
each of them the essential sensing data-generator features.

\begin{table*}[t]
    \centering
    \caption{Summary of skin conductivity equipment.}
    \label{table:Skin}
    \begin{tabular}{|c|c|c|c|c|c|c|c|c|}
        \hline
        \textbf{Producer} & \textbf{Model} & \textbf{Grade} & \textbf{Mobility} & \textbf{Conductivity range} & \textbf{Channels} & \textbf{Sampling rate/channel} & \textbf{Encoding} & \textbf{Reference} \\
        & & & & \textbf{[S]} & & \textbf{[Hz]} & \textbf{[bits/sample]} & \\ \hline

        Neulog & NUL-217 & clinical & wearable & 0-10 $\mu$S & 1 & 100 & 16 & \cite{Neulog2016} \\ \hline
        Shimmersensing & Shimmer3 GSR+ & clinical & wearable & 0.2-100 $\mu$S & 1 & N/A & N/A & \cite{Shimmersensing2016} \\ \hline
        Mindmedia & SC Sensor & research & wearable & 0.1-1000 $\mu$S & 1 & N/A & N/A & \cite{MindmediaGSR} \\ \hline
        TMSI & GSR sensor & research & mobile & N/A & 1 & 100 & N/A & \cite{TMSI2016} \\ \hline
        Mindfield & eSense & hobby & wearable & N/A & 1 & 10 & 18 & \cite{Mindfield2016} \\ \hline
        TMSI & GSR sensor & research & mobile & N/A & 1 & 100 & N/A & \cite{TMSI2016} \\ \hline

    \end{tabular}
\end{table*}

\item \textbf{Movement, positioning and gait sensors:} Recent
advances in miniaturisation of inertial sensors led to an increase
in monitoring of physical movement and positioning of humans in
relation to different contexts. In medical sciences inertial
sensors can measure and monitor neuro-degenerative conditions such
as Parkinson's disease, patient's fall due to seizures or cardiac
arrest, or patient's movement during physical rehabilitation
programs. Another extensive use of inertial sensors is in sports
and entertainment for monitoring and improving certain physical
activities and the relevant cognitive-motor skills. Last, but not
least, in recent years there has been a surge of studies which
employed wearable inertial sensing for monitoring and
investigation of social activity and behaviour.

The various mobile, portable or wearable inertial sensors for
human activity follow the same architecture and technology as the
general inertial sensors we discussed earlier in the paper,
consisting of combinations of accelerometers and gyroscopes for
measuring acceleration along a sensitive axis and over a
particular range of frequencies. Most of the sensing systems
reported in the literature are based on piezoelectric,
piezoresistive, variable capacitance, or MEMS transducers, which
use variations of the classic inertial principle of operation: a
core element (i.e. a mass) that responds to acceleration by
proportional stretch or compression of another spring-like
element. An thorough review on inertial sensors for human activity
monitoring, with a focus on wearability can be found in
\cite{Sprager2015}.

Table~\ref{table:Movement} summarises the most popular
movement/activity trackers available to date, and list for each of
them the essential elements of the fused activity sensing system.

\begin{table*}[t]
    \centering
    \caption{Summary of movement, positioning and gait sensing equipment. Legend: A - accelerometer, G - gyroscope}
    \label{table:Movement}
    \begin{tabular}{|c|c|c|c|c|c|c|}
        \hline
        \textbf{Producer} & \textbf{Model} & \textbf{Mobility} & \textbf{Format} & \textbf{Inertial sensors} & \textbf{Other bio-sensors} & \textbf{Reference} \\ \hline

        Angelsensor & Classic/M1 & wearable & wristband & A, G & HR, skin temperature, oximeter & \cite{Angelsensor2016} \\ \hline
        Apple & Watch & wearable & watch & A & HR & \cite{Apple2016} \\ \hline
        Basis & BasisPeak & wearable & watch & 3 axis A & HR, GSR, skin temperature & \cite{Mybasis2016} \\ \hline
        Empatica & E series & wearable & wristband & 3 axis A & HR, GSR, skin temperature &
        \cite{EmpaticaE} \\ \hline
        Empatica & E series & wearable & wristband & 3 axis A, G & HR, GSR, skin temperature & \cite{EmpaticaEmbrace} \\ \hline
        Fitbit & Zip & wearable & pendant & 3 axis A & - & \cite{FitbitZip} \\ \hline
        Fitbit & One/Flex/Charge/Alta & wearable & wristband & unspecified motion sensor & HR, simplified ECG & \cite{FitbitWristbands} \\ \hline
        Fitbit & Blaze/Surge & wearable & watch & unspecified motion sensor & HR, simplified ECG & \cite{FitbitWatch} \\ \hline
        Fitbug & Orb & wearable & watch & 3 axis A & - & \cite{Fitbug2016} \\ \hline
        Garmin & VivoFit/VivoSmart series & wearable & wristband & unspecified motion sensor & - & \cite{Garmin2016} \\ \hline
        Garmin & VivoActive series & wearable & watch & unspecified motion sensor & - & \cite{Garmin2016} \\ \hline
        Google & Android-Wear & wearable & watch & market dependant & market dependant & \cite{Google2016} \\ \hline
        iHealth & Edge & wearable & wristband & unspecified motion sensor & - & \cite{iHealth} \\ \hline
        Jawbone & UP series & wearable & wristband & 3 axis A & HR, GSR, respiration & \cite{Jawbone2016} \\ \hline
        LG & Lifeband Touch series & wearable & wristband & unspecified motion sensor & - & \cite{LGLifeband} \\ \hline
        LG & watch series & wearable & watch & unspecified motion sensor & - & \cite{LGWatch} \\ \hline
        Medisana & ViFit series & wearable & wristband & unspecified motion sensor & - & \cite{Medisana2016} \\ \hline
        Microsoft & Band & wearable & wristband & 3 axis A, 3 axis G & HR, GSR, skin temperature & \cite{MicrosoftBand} \\ \hline
        Misfit & Shine/Ray series &  wearable & watch & 3 axis A & - & \cite{Misfit2016} \\ \hline
        Misfit & Flash & wearable & watch & 3 axis A & - & \cite{Misfit2016} \\ \hline
        Misfit & Ray &  wearable & pendant & 3 axis A & - & \cite{Misfit2016} \\ \hline
        Misfit & Link &  wearable & button & 3 axis A & - & \cite{Misfit2016} \\ \hline
        Nike & Fuelband & wearable & watch & 3 axis A & - & \cite{Fuelband2016} \\ \hline
        Pebble & SmartWatch series & wearable & watch & 3 axis A & - & \cite{Pebble2016} \\ \hline
        Razer & Nabu/NabuX & wearable & wristband & 3 axis A & - & \cite{RazeroneNabu} \\ \hline
        Razer & Nabu Watch & wearable & watch & 3 axis A & - & \cite{RazeroneWatch} \\ \hline
        Salutron & LifeTrak series & wearable & watch & 3 axis A & HR, ECG & \cite{Salutron2016} \\ \hline
        Samsung & GearS series & wearable & watch & A, G & HR & \cite{SamsungGearS} \\ \hline
        Samsung & GearFit series & wearable & wristband & A, G & HR & \cite{SamsungGearFit} \\ \hline
        Sensoria & SensoriaFitness & wearable & smart sock & 3 axis A & 3 pressure sensors & \cite{Sensoriafitness2016} \\ \hline
        Sony & Smartband series & wearable & wristband & A & - & \cite{Sony2016} \\ \hline
        Sony & Smartwatch series & wearable & watch & 3 axis A, G & - & \cite{Sony2016} \\ \hline
        Striiv & Fusion series & wearable & wristband/watch & 3D accelerometer & - & \cite{Striiv2016} \\ \hline
        Withings & Pulse O2 & wearable & pendant & 3 axis A & heart rate, oximeter & \cite{WithingsO2} \\ \hline
        Withings & Go & wearable & pendant & 3 axis A & - & \cite{WithingsGo} \\ \hline
    \end{tabular}
\end{table*}

\end{itemize}

\subsubsection{Sensing machine agent activity}

Surrounding technological systems are also one of the vital
sources of data for autonomous systems. An autonomous unmanned
road vehicle, for example, will gather data from various machine
agents (i.e. peer-technology sources) that can be internal or
external in relation to the vehicle. Internally, numerous sensors
monitor the mechanical sub-systems involved in the normal
operation of the vehicle, such as propulsion (e.g. crankshaft,
camshaft, torque, oxygen, temperature sensors, etc.), breaking
(ABS - Anti-lock Break System, EBD - Electronic Break
Distribution), safety (impact, light, rain, proximity sensors),
etc. Externally, numerous other sensors monitor the interaction of
the vehicle with the environment, such as GPS for positioning,
long and short range radar for obstacle detection, cameras for
vision, etc. The situation becomes even more complex when speaking
about the operation of an aircraft, or about the operation of a
whole aircraft fleet, or an even more complex example, the
operation of an autonomous air traffic control system, which
actually handles an entire airspace. The resultant sensing
mechanism is an overall equivalent sensor of very high complexity,
which provides a fused global image of the data coming from the
numerous internal and external sources.

In the previous section we discussed human activity sensing by
considering the main categories of biophysical activity sensors
separately. That discussion was pertinent because humans are
identical from the perspective of the activity data they produce,
thus autonomous systems need to embed a unique set of sensors in
order to interact with different humans. However, in the case of
sensing the activity of technical systems, there is a variety of
sources the autonomous system may encounter during its operation.
Thus, it becomes pertinent to discuss fusion sensing rather than
the separate types of sensors. Sensor fusion \cite{Elmenreich2007}
is the process of combining the sensory data coming from disparate
sources in a way that (1) reduces the uncertainty resulting from
separate treatment of the sources and/or (2) unveils properties of
the sensed environment that were not obvious when using separate
sources. Sensor fusion is receiving increasing attention from the
research community, as the sensor technologies improve, together
with the subsequent autonomous systems. Several thorough
discussions can be found in the literature, proposing fusion
methods with application in various fields of activity, such as
control systems \cite{Khamseh2016}, thermal engineering
\cite{Jiang2016}, wearable robotics \cite{Novak2015}, or 3D
computer vision \cite{Tansky2014}.

Over time, numerous sensor fusion architectures have been proposed
and categorised differently by different studies. From a project
management perspective, in \cite{Elmenreich2007}, the author
describes abstract, generic and rigid architectures. The abstract
ones constitute as ways to describe or explain the operation of a
fusion system without guiding the engineers toward the actual
implementation. Their importance lies in that they provide the
initial understanding of the fusion problem. Generic architectures
describe how to implement the system, but without specifying the
technology (i.e. operating system, software and hardware,
communication system). They are important for providing the
engineers with complete know-how, while still leaving them the
freedom to adapt their solutions to specific problems. Rigid
architectures provide detailed design and implementation of the
fusion system, guiding the engineers step-by-step in a strict
manner. They are important when sensing systems are embedded in
critical infrastructures or safety critical applications, where
rework or re-implementations of the systems are not only
unnecessary, but actually strongly avoided.

Another classification proposed in the literature
\cite{Durrant-White2002,Mirza2008,Azimirad2015} takes a conceptual
approach and discusses four main categories - centralised,
hierarchical, distributed (or autonomous), and decentralised -
from which various hybrid architectures can be further generated
\cite{Mirza2008}. This classification is perhaps the most popular
for both researchers and practitioners, becoming in time a
fundamental part of the undergraduate control engineering
curriculum \cite{Durrant-White2002}.

In centralised fusion architectures raw sensor data are
transmitted to a central site, where they are aggregated to
generate the information needed to produce a single fused
description of the state/dynamics of the environment. In general,
centralised fusion needs high processing power and data buses with
high band-width in order to handle the high amount of raw data
generated by sensors; thus, scaling up to complex systems may
became a problem.

In hierarchical fusion, the lowest level processing elements
transmit information upwards, through successive levels, where the
information is combined and refined at each level, until it
reaches the top level, where the full view of the state/dynamics
of the environment is made available. Hierarchical fusion involves
a certain processing capability on each level, and thus the load
on the centralised (highest level) processor is reduced. In the
same time the need of communication is reduced, since transmission
of data/information is strictly controlled and limited to adjacent
layers. However, while useful in general, this can be also a
disadvantage due to the inability of data/information sources to
communicate over more than one layer.Another major disadvantage of
hierarchical fusion is that changes in a certain layer imply
changes in all related sub-layers; this usually prevents the
sensing system to be extended to incorporate more sensors.

Distributed fusion was meant to mitigate the drawbacks of
centralised and hierarchical fusion, by providing autonomy to
various sensing entities. In there architectures expert sensing
agents with own processing capabilities make available to the
system the output of their operation through a communication
platform, often known as a ``blackboard'' from the architecture
with the same name \cite{Mirza2008}. The obvious advantage of this
approach is the distributed processing, with no central processing
unit or layer. However, problems may arise due to the use of
common communication/memory resource, i.e. the blackboard. Another
disadvantage may be a potential loss in the accuracy of the global
picture of the environment, due to the fact the picture is now
dependant on simple agents with various processing capabilities.

Decentralised data fusion abandon completely all commonality
aspects. A decentralised system consists of a network of sensor
nodes, each with its own processing facility, and no central or
common communication facility. In this case, fusion happens in
each node, locally, by the sensor unit using local observations
and information communicated from neighbouring nodes. Advantages
of the decentralised approach are high scalability and modularity,
where dynamic changes of the sensor number and structure are
easily supported. Thus, the resultant sensing systems are highly
resilient to loss of nodes. However, due to the lack of global
communication or knowledge, the accuracy of the fused picture of
the environment is highly dependant on the network features, i.e.
topological features like centralities, entropy and other
statistical metrics.

Numerous architectures have been proposed over time, from as early
as 1960s to present days, for each of the four (plus the hybrid
approaches). The interested reader can find thorough reviews of
the general fusion architectures in a number of studies from
different historical periods
\cite{Dasarathy1997,Elmenreich2007,Mirza2008,Azimirad2015}.
However, it is outside of the scope of this study to provide
insights in the individual architectures, since our purpose was to
bring into attention the complexity that a sensor-based autonomous
system must handle, in terms of the amount and variety of sensory
data coming from the surrounding technological systems. Thus, we
emphasize once more on the idea that big sensory data can be in
the same time an enabling and a disabling factor in the operation
of the autonomous systems.

 \subsection{Computers}

Once data-driven and model-driven information become available,
processing starts. Be it in the form of signal processing, feature
selection, clustering, classification, or any form of processing
that is needed to make a judgement, the machine needs to rely on
its processing abilities to process these information. Processing
today extends beyond the classic Von Neuman computer architecture
to other forms of processing.

In the most general understanding, a computer is an entity that
conveniently alters a set of inputs to generate a set of desired
outputs. Since digital computers started to be used on large
scale, and became indispensable to modern life, the general
perception of a computer narrowed to its digital incarnation,
where inputs are reduced to binary data and alteration towards the
desired output is made through binary calculation. However,
computing was and is not limited to digital manipulation of binary
input data, since numerous entities exist that can perform the
computing analogically if talking about machine entities (e.g.
mechanical, electrical, optical, electronic) or biologically if
talking about natural entities, such as the neural sensory
processing in animals with a central nervous system
\cite{MacLennan2004,MacLennan2009}. Given that in this study we
discuss autonomous systems considering the big sensory data they
have to process during their normal operation, it is natural to
bring into attention mostly the digital computing machines
available to date, with a focus on their ability to deal with
these data. However, we consider useful to discuss the other two
categories, analog computing and neural-inspired computing, in
order to offer a more comprehensive and integrated view on the
ways that are open to autonomous systems.

\subsubsection{Analog computing}

Analog computers use the continuously changeable aspects of a
physical phenomenon of interest to alter the inputs towards the
desirable output. Virtually, any physical process that models some
computation can be seen as an analog computer, including finding
the shortest path between two points by stretching a string
between them. They have been heavily used in scientific and
industrial applications in classic automation and control systems
of the pre-digital era, when digital computers of the time lacked
sufficient performance. However, they have been less and less
used, becoming eventually obsolete as the digital computers came
into place and became more powerful and versatile. There are still
domains where they are still in use in some specific applications,
where classic automation is required due to operation costs and
environmental conditions. Extensive historical views on analog
computers can be found in \cite{Small1993}, and also in a special
issue of IEEE Control Systems dedicated to the history of analog
computers, from which we cite here the editorial article
\cite{Lundberg2005} signed by Lundberg.

Examples of analog computers can go from the ancient nomographs
and sextants, to slide rules, and further to complex contemporary
automation and control systems such as navigation systems, analog
auto-pilot systems for vehicle navigation, military weapon system
controllers, analog signal processors, etc. In essence, virtually
all industrial process control systems in the pre-digital era used
analog computers, known at the time as controllers, to
automatically regulate temperature, flow, pressure, or other
process variables and parameters. The technology of these
controllers ranged from purely mechanical systems to emulation of
physical phenomena through electronic analog components, and later
through analog integrated circuits.

The strongest point the analog computers make when compared to
their digital counterparts is the computation speed; i.e. the
analog computers operate in real-time once they have been set up.
The computation in analog computers is not explicit like in
digital computers, but rather implicit, inherently contained in
the materials or phenomena they use as computation mean; thus the
delivery of the output based on the input is virtually
instantaneous. For example, an integrator can be built as an
electric circuit by conveniently placing a resistor and a
capacitor in a feedback loop, so that the output delivers a
steadily changing voltage for a constant input voltage.

A major disadvantage is that analog computers are designed for
performing calculation in particular contexts outside of which
they cannot function properly. Their setup involves choosing (1)
the scale and limits for the inputs and outputs given the material
or phenomena used for computation, (2) the set of pre-operational
(initial) and operational conditions, and (3) the mechanism or
physical structure (i.e. interconnection of computing elements) of
the computation in order to assure proper solving of the given
problem. When launched into operation variables cannot exceed
computer's predefined operational conditions. For example, if the
above mentioned integrator is to be used in humidity and
temperature conditions that exceed the limits established during
design stage based on the material/electrical tolerances indicated
by manufacturer for the component elements (resistor, capacitor
etc.), then, unacceptable calculation errors may occur. Also, if
the integrator is to be used as a differentiator, this can be done
only by changing the physical arrangement of the resistor and
capacitor in the feedback loop, together with their values. In
other words, a different circuit - that is, a different analog
computer - is needed.

Analog computers can be found or imagined virtually anywhere and
in anything in nature, however, the literature mostly speaks about
the mechanical and electronic versions, due to the fact they
account for the majority of the modern time achievements in both
industry and research in pre-digital era. Both categories stand
out due to their theoretical importance at the time or due to the
popularity they gained in practical industrial applications. In
the following we briefly discuss them in order to create a
complete image of the analog computing field.

\begin{itemize}

\item \textbf{Mechanical analog computers:} Mechanical analog
computers used over the years were able to implement the usual
arithmetic, trigonometric, and geometric operations, and also
complex analytic and algebraic transformations or arbitrary
functions by conveniently combining various mechanical elements,
such as rotating shafts, discs and gears, racks and pinions,
cables and pulleys, springs, differential mechanisms, etc. For
example, addition and subtraction have been implemented using
precision bevel-gear differentials. Integration was implemented
with a rotating disk (integrator variable) and a pickoff wheel
positioned on the disc at a radius proportional to the integrated
variable. Also, coordinate conversion from polar to rectangular
was implemented with two coaxial discs and a sliding block with a
pin on it. While these examples are important for analog
computers, they are certainly not exhaustive. Extensive reviews of
these computers can be found in the literature of the 1960s and
1970s \cite{Rogers1960}, and more recently in
\cite{MacLennan2009}.

\item \textbf{Electronic analog computers:} Electronic analog
computers emerged as a result of the similarities between the
equation-based mathematical models of mechanical components, and
those of the electrical components. Complex and heavyweight
mechanical computational devices were gradually replaced with
their electrical equivalents which could be constructed with
operational amplifiers and other linear (resistors, capacitors,
inductors) and non-linear (diodes, transistors) passive
components. The resultant electronic analog computers were less
expensive to construct, safer, and easier to modify for new
problems in comparison with the mechanical counterparts.
Gradually, more and more technological and natural phenomena could
be represented through analogy by using electronic circuits, which
could easily perform a wide variety of simulations. For example,
an application would be to integrate a signal representing water
flow, producing an output signal representing total quantity of
water that has passed through a flow-meter. Voltage can be the
analogy of water pressure and electric current the analogy of
flow-rate in terms of volume/second. Then, the integrator can
provide the total accumulated volume of liquid, using an input
current proportional to the flow rate.

Electronic analog computers contain may contain, depending on
their purpose, tens, hundreds or thousands of operational
amplifiers combined with relevant passive components in various
feedback schemes, in order to implement mathematical operations
and functions. The core operations and functions are: addition,
subtraction and comparison; differentiation and integration;
multiplication, division and inversion; exponentiation and
logarithm. More complex arbitrary non-linear functions have been
also implemented, however, they had limited precision (typically
three to four digits) and were difficult to implement, requiring
laborious and complicated feedback loops with combinations of
linear and non-linear components in the feedback loops of
operational amplifiers. An in-depth discussion on electronic
analog computers can be found in \cite{Small2001}.

The major disadvantages of the electrical analogy are (1) the
limited dynamic range, i.e. electronic components can represent
limited variation ranges of the variables they emulate (compared
for example with the virtually infinite dynamic range of the
floating-point representation used by digital computers), and (2)
the noise levels introduced by the electronic components and
circuits. In general, electronic analog computers are limited by
non-ideal effects induced by the undesired deviations of the core
variables of the electric signals: DC and AC amplitudes,
frequency, and phase due to material and environmental limitations
\cite{Wambacq1998}. The limits of range on these characteristics
limit analog computers. Some of these limits include the
operational amplifier offset, finite gain, and frequency response,
noise floor, non-linearities, temperature coefficient, and
parasitic effects within semiconductor devices.

\item \textbf{Revival of analog computing:} Analog computation
certainly went into an eclipse in the last few decades, due to the
rise of digital computers, however recently more and more authors
acknowledge that analog computation is inherent to most of the
existing natural systems \cite{Daniel2013}. In
\cite{MacLennan2004} the author even note that some digital
computing paradigms are actually unsuited and unnatural (pun
intended) to natural computation. Thus, recently new iterations of
analog computers have been proposed, especially in their
electronic version.

Recent development of semiconductors and very large scale
integration went mostly in the direction of integrating digital
circuits. However, recently, several analog VLSI circuits have
been proposed, implementing energy efficient and instantaneous
complex mathematical calculation. Cowan and colleagues
\cite{Cowan2005} proposed a single-chip VLSI analog computer with
80 integrators and 336 other programmable linear and nonlinear
circuits manufactured in $250 nm$ CMOS technology. The chip was
intended for use as an analog mathematical coprocessor for general
purpose digital processors, and was rated by authors with a
consumption of only $300 mW$. Later, Schell and Tsividis
\cite{Schell2008} proposed a hybrid signal processor where fixed
sampling and clock rates of a conventional DSP were eliminated to
create a clock-less digital signal processor. The processing unit
was manufactured in $90 nm$ CMOS technology and operated in
continuous time with a power source of 1 Volt, offering no
aliasing, and very fast transient performance in power electronics
due to lack of sampling. Recently Guo and colleagues
\cite{Guo2015} demonstrated a continuous-time hybrid computing
unit in $65 nm$ CMOS technology, capable of solving nonlinear
differential equations up to 4th order, and scalable to higher
orders. The authors rate the energy efficiency of the processor at
$12 mW$, and demonstrate the use of the unit in a low-power
cyber-physical systems application, where the problem is solved in
$0.84 \mu s$.

\end{itemize}

\subsubsection{Digital computing}

In general, analog computers, analog signals or other analog
processes implicitly involve an \textit{analogy} or a systematic
relation with the physical processes they model or simulate. They
have inherent advantages, as discussed in previous section, such
as speed and low energy consumption, given by the fact they
operate on continuous representations of real processes by using
continuous processes \textit{analog} to real ones. However, the
\textit{analogy} itself also creates one of the most important
drawbacks, that is the lack of flexibility, since a signal that
was found to be \textit{analog} to a physical process typically
cannot be used with ease for another process. Digital computers,
on the opposite, operate on discrete representations of real
processes by means of discrete steps. Thus, real processes are
represented by strings of binary symbols that do not have an
explicit or direct physical relationship to the modelled
processes. In \cite{MacLennan2009} the author notes that in a
digital computer the relationship, i.e. \textit{analogy}, is more
abstract than simple proportionality. This extra abstraction steps
made the digital computers extremely versatile, in that the
discrete modelling and operation allow their use in a large
variety of applications, virtually without any hardware change.
With this important advantage, doubled by a rapid increase in
speed and energy efficiency, the digital computing eclipsed the
analog ones, despite a number of disadvantages. Some of these
advantages, which we mention here for completeness reasons, are as
follows: (1) sampled signals used in digital computing suffer from
aliasing effect, which leads to power dissipation on superior
harmonics of the digitised signals and to the need of precise
filtering; (2) digital signals may present important undesired
behaviour in the transient domain; (3) discrete representation can
lead to latency when data are transferred between various internal
components of the processing architecture.

Today, digital computers reached an unprecedented diversity, with
their performance measured with respect to certain attributes that
make them fit-for-purpose, rather than a unique performance
metric. In general, digital computing is subject to a trade-off in
which multiple aspects are involved, such as processing-related
features (e.g. energy consumption, memory capacity, throughout,
latency, etc.), operational features (e.g. size, weight,
reliability, expandability, etc.) on top of which there are of
course the financial costs of acquisition and operation. Thus, the
overall performance of a digital computer can be measured using
numerous metrics depending on its purpose. For example, a computer
can be CPU bound - heavily used for raw computation, input-output
(I/O) bound - mainly used as a server, or memory bound - mainly
used for memory intensive tasks such as video editing. From the
point of view of autonomous systems though, two aspects are
essential: raw processing power and energy efficiency. We
discussed earlier in the sensors section how portability and
miniaturisation are the engines that propelled the sensors field
to nowadays advances, which are unprecedented in the history of
human kind. Thus, the autonomous systems making use of sensors
must fall in the same philosophy of fast processing of data versus
high portability.

Processing power was first associated with processor's clock rate,
that is, the frequency (number of cycles per second) of the main
clock of the CPU. However, this turned misleading as digital
computing evolved and processors became able to execute more and
more instructions at every processor cycle. Thus, a higher clocked
processor may have a lower throughput than a lower clocked
processor if the latter can execute significantly more
instructions per cycle. Thus, another measure of the processing
speed was the number of instructions per cycle, which in
conjunction with processor's clock rate could give a better
estimate of the speed with at which a certain task can be
processed. However, this was also subject to ambiguities since
different processor architectures can embed different sets of
instructions, and/or can implement similar instructions
differently. Perhaps the most accepted speed metric for modern
processors is related to the amount of work done per unit time,
i.e. the throughout, and is measured as the number of floating
point operations per second (FLOPS).

Energy efficiency is the other vital aspect in the operation of
autonomous systems, mostly for portability reasons, especially
because the usual perception of an autonomous system involves
embodiment and mobility. Disembodied autonomous systems can be
brought into discussion, like autonomous decision-making systems
hosted on mainframes of fixed computing facilities, in which case
the energy consumption may be important from a cost perspective.
However, in a sensing-based autonomous system, mobile embodiment
is the pertinent form of existence, which leads to high importance
of the energy efficiency offered by the processing units.

\begin{itemize}

\item \textbf{Classic processing-storage digital computing:}
Computers in this category generally refer to architectures in
which memory is conceptually separated from the processing unit,
and data to be processed are permanently exchanged between the
two. In general, the processing-storage architectures describe
digital computers consisting of a processing unit, memory units
that store programs (instructions) and data, input-output
mechanisms, and external mass storage. Several architectures have
been proposed over time, which differ through the way the exchange
takes place within the component elements of the architecture. Of
these, \textit{von Neumann}, \textit{Harvard} and
\textit{modified-Harvard} architectures are those that provided
the majority of the digital computers available to date
\cite{Singh2014}.

The von Neumann architecture proposed a design in which both data
and instructions are stored in a unique memory, and thus they
share one data bus for communication with the processing unit. The
architecture was simple, through the use of only one data bus and
one memory, but also introduced a major disadvantage, known as
\textit{von Neumann bottleneck}. The bottleneck consisted in the
impossibility to fetch instructions and operate on data in the
same time due to the shared bus that could only handle one type of
memory items at a time. This led to limitations of the overall
throughput (processing power), especially in data intensive
applications, due to the fact the CPU was continually forced to
wait for data to be transferred to or from memory.

Several methods to overcome this bottleneck have been proposed
over time, and included sole or combine use of: a cache between
CPU and memory, separate caches for data and instructions between
CPU and memory, separate access paths (bus) for data and
instructions, on-chip single- or multi-layer CPU cache, pipeline
mechanisms with prediction algorithms for data/instruction
transfer, etc.

The Harvard architecture was one of the most successful
improvements proposed for overcoming von Neumann bottleneck. The
architecture introduces a design with physically separate storage
and transfer for instructions and data, thus the design features
dedicated data memory and instruction memory, as well as dedicated
data bus and instruction bus. Thus, in computers build on Harvard
architecture, the CPU can read an instruction and operate a data
transfer at the same time, leading to higher processing speed
without using memory cache. In addition, since data and
instructions are completely separated (storage and transfer path)
their features and implementation can be different, such as
different word size, different timing, different implementation
technology, or different memory address structure, as well as
different data transfer standards.

The main use of Harvard design, is in applications where the
energy saving and implementation costs, so that processor cache
must be avoided, while still keeping a high processing speed. In
signal processing, the digital signal processors (DSP) generally
execute small, highly optimised processing algorithms, hence their
memory usage is well determined, so that memory cache is not
needed. In the same time, DSP applications need high parallelism,
so multiple address spaces must be used, that usually need to be
addressed in parallel. Thus, some DSPs feature multiple data
memories in distinct address spaces to facilitate paralel
processing approaches like Very-Long-Instruction-Word (VLIW) and
Single-Instruction-Multiple-Data (SIMD). An example is the Texas
Instruments TMS320 C55x processor family, which includes five
parallel data buses two for writing and three for reading, and one
instruction bus. Micro-controllers are another category of
computing units implemented in Harvard architecture. They are
typically used in embedded systems, where the applications need
small amounts of program  (instructions) and data memory, but need
high speed processing with low energy consumption. Thus, designs
with concurrent instruction and data access and no cache are
preferred by most manufacturers, such as Atmel's AVR family or
Microchip's PIC family.

However, while providing high processing speed and low energy
consumption for a range of applications like RISC or embedded
systems, the architecture was not appropriate for full-sized
modern processors used in general-purpose computing, such as
personal computers and laptops, and more recently tablets and
smart-phones. This happened especially due to the unbalanced
development of various elements of the computing architecture,
where for example the speed of CPU, memory, bus and external
storage experienced very different growth pace.

Thus, various improvements of Harvard architecture have been
proposed, all known under the umbrella of modified Harvard design.
In essence the modified versions are still Harvard architectures
through that the CPU is able to access data and instructions
separately, but in the same time they resemble von Neumann design
by storing both in a unique memory. The use of
\textit{split-cache} system is the most common modification, used
in most modern general-use processor families from INTEL, AMD,
ARM, Motorola, IBM, SUN, etc. In this design, separate instruction
and data caches are used by CPU to access a common address space.
When the CPU executes from caches the computer falls onto the
standard Harvard design, and when it accesses the common memory it
falls onto the von Neumann design.

\item \textbf{Highly parallel processing on GPUs:} A significant
step towards increasing processing speed was made during the mid
2000s, when general-purpose computing tasks started to be ported
from CPUs onto graphic processing units (GPU)
\cite{Brodtkorb2013,Khatter2014}. GPUs are specialised processing
units designed to accelerate the creation of images in a frame
buffer (that is further output-ed to a display) through a
multi-core structure that is able to process large blocks of
visual data in parallel. GPUs operate in general at lower clock
frequencies compared to CPUs, but embed very high numbers of
simple cores working in parallel, overpassing overall the
capabilities of even the fastest multi-core CPUs available today.
However, the limitation of GPUs was that their simple cores were
designed for handling graphics data only. Early GPUs embedded
specialised cores were used for texture mapping and polygon
rendering, with later additions that included various geometric
calculations, e.g. rotation/translation of vertices into different
coordinate systems. More recent GPUs added programmable shaders
which for manipulating vertices and textures, oversampling and
interpolation techniques for anti-aliasing, and very
high-precision color spaces. Today, typical specialised cores in
modern GPUs are pixel shaders and vertex shaders (recently
combined in ``unified shader architecture''), texture mapping
units, and render output units.

Since most graphical computations involve vector and matrix
manipulation, researchers and practitioners have increasingly
studied the use of GPUs for non-graphical calculations. Since the
graphic cores were unable to process general-use data, the usual
CPU applications had to be transformed into graphical forms in
order to be processed by GPUs. Thus, early attempts to perform
general-purpose computing on GPUs focused on reformulating at
conceptual level the general computation problems in terms of
graphics primitives, in order to be compatible with OpenGL or
DirectX \cite{Microsoft2016}, the two major APIs available at the
time. This difficult translation was later abandoned with the
emergence of the GPU computing ecosystems (API + SDK), such as the
proprietary CUDA from nVidia \cite{nVidia2016}, which was the
first API to allow the transfer of C, C++ and Fortran code (and
later Matlab, Python and C$\#$) straight to GPU, with no assembly
language required. Later, in 2007, AMD-ATI released its own
proprietary ecosystem Stream, which followed the ATI developed
Close-To-Metal and further evolved into the recent Accelerated
Parallel Processing (APP) ecosystem that offers support for newest
AMD's Heterogeneous System Architecture, in which CPU and GPU
cores are designed to work together in a single accelerated
processing unit (APU) \cite{AMD2016}. Recently,
hardware-independent platforms have been released, such as the
proprietary DirectCompute from Microsoft as part of DirectX API
\cite{Microsoft2016}, and the open-source OpenCL
\cite{Khronos2016}. Through these, the general-purpose
applications can be run through and benefit of GPU speed without
requiring conversion of the data into graphic form.

CUDA and Stream/APP ecosystems remained bond to nVidia and AMD
GPUs, respectively. nVidia GPUs starting with GeForce 8 series
were by default hardware enabled for CUDA, while AMD's GPUs
starting with Radeon500 family were by default hardware enabled
for Stream (and later for APP). Recently, the open source
platform-independent OpenCL, developed by Khronos Group
\cite{Khronos2016} became popular for that it allows development
of code for both GPUs and CPUs, and received support from all
major chip manufacturers, Intel, AMD, Nvidia, and ARM, whose most
recent chips are OpenCL-enabled.

\item \textbf{Reconfigurable processing:} Apart from the
processing power and energy consumption, another aspect becomes
more and more important in the context of sensor-enabled
autonomous systems. Since the nature of these systems, as we
explained in earlier in the paper, is to operate in dynamic
environments, it becomes important to be able to adapt not only
the decision making, but also the processing itself to the task of
interest. However, while this is possible on general use computers
through software, it is not possible at lower hardware level.
Thus, an increased interest have been shown for the use of FPGAs
(Field-Programable Gate Array), which by design have the potential
to be reprogrammed at runtime, being in essence computing units
that can reconfigure themselves to suit the task of interest
\cite{Cumplido2015}. Typically, FPGAs contain arrays of
programmable logic blocks, and a hierarchy of reconfigurable
interconnects that allow the blocks to be connected conveniently
to perform a variety of operations, from simple logical operations
like XOR, to complex combinatorial functions, mathematical
functions or even emulate whole processors. Historically, they
have been used as ``breadboards'' for low cost prototyping, in
order to avoid the high non-recurring engineering costs of ASIC
(Application-Specific Integrated Circuit) development. Thus, chip
manufacturers could develop their hardware on standard FPGAs, and
then manufacture at large scale their final version as an ASIC
that could not be modified once the final commitment to a design
has been done. For the same reason, FPGAs were also preferred for
designing and manufacturing of chips, in applications where the
volume of production was small and the development of ASICS was
not worthy at all. Initial FPGAs were bond to these purposes also
because of their low performance, combined with large physical
dimension and high energy consumption \cite{Hassan2010}. However,
with the recent advances in semiconductor technology, transistor
gates and memories, the latest FPGAs can rival not only with their
ASIC counterparts, but also with full-sized processors in multiple
respects, such as speed, dimension, energy efficiency and cost,
while having the advantage of reconfigurability. Recent FPGAs
began to take over larger and larger functions to the point where
they are even able to implement so-called ``soft processors'' that
emulate the operation of powerful computers, either in ASIC domain
or in general-computing domain. FPGA-based soft processors can be
seen in various reconfigurable System-On-Chip (SoC) solutions,
which incorporate in the FPGA fabric a variety of elements, such
as ARM cores, networking modules, radio modules, DAC/ADC
converters, and other peripherals. Examples of these FPGA are
those from the major chip manufacturers, such as Microsemi
SmartFusion series, Xilinx Zynq series, or Altera Arria series. A
detailed review of the high speed FPGA SoC and soft processors can
be found in \cite{Lysecky2005}.

Due to these technological advances, FPGAs are used today in
complex applications that include digital signal processing,
software-defined radio, medical imaging, computer vision, speech
recognition, cryptography, bio-informatics, computer hardware
emulation, radio astronomy, metal detection and a growing range of
other areas. Another recent use of FPGAs is hardware acceleration,
where FPGAs are used as coprocessors, to assist a general-use
processors in acceleration of certain parts a task or algorithm
\cite{Ekas2003}. Recent use of FPGA coprocessors was announced by
Microsoft and Google, who started to use FPGAs to accelerate
high-performance, computationally intensive data centres that
operate their search engines, i.e. Bing and Google respectively.

\end{itemize}

\subsubsection{Neuromorphic computing}

Neuromorphic engineering, also known as neuromorphic computing,
has its roots in the analogic nature of natural biological
computation and is based on the transfer of the morphology of
individual into electronic circuits. In essence, the resultant
computing architectures and implementations are analog computers
which contain circuits that mimic the electrical activity of
neuro-biological nervous system, that is, described by the spiking
neuron models. From the point of view of a sensor-based autonomous
system, neuro-morphic computing seems like the obvious way to go,
since the sensor-based operation is the essence of life, and thus
any existing digital artificial sensor-based system, no matter how
fast or sophisticated it is, is still far from what biological
sensory processing can offer.

Most of the efforts in neuromorphic computing concentrated on the
formal spiking models, with some notable instantiations of the
theoretical models proposed in the last decade.

In 2006, researchers at Georgia Institute of Technology published
the details of a field programmable neural array
\cite{Farquhar2006}. The chip was the first in a line of many
increasingly complex arrays of floating gate transistors that
allowed programmability of charge on the gates of MOSFETs to model
the channel-ion characteristics of neurons in the brain and was
one of the first cases of a silicon programmable array of neurons.
The authors describe the chip as an analog circuit capable of
accurately emulating large complex cells, or multiple less complex
ones. They note its analogy to an FPGA in which the binary
transistor gates were replaced by biologically relevant circuit
components including active channels, dendrites, and synapses. A
routing capability was added to these elements in order to
transfer outputs from cells to individual synapses.

In 2012, Spintronic Rand Purdue University \cite{Sengupta2015}
announced a neuromorphic chip that used lateral spin valves and
memristors. Their design was a hybrid structure in which a
programming current passing through heavy metal generates
spin-orbit torque to modulate the spintronic device conductance,
creating a nanoelectronic synapse. The spintronic synapse was then
interfaced with CMOS neurons operating in transistor subthreshold
region in order to form a network of spiking neurons. The authors
claim that the architecture can emulate the operation of
biological neurons and reproduce various brain-like processing
abilities with significantly lower energy consumption than
conventional digital chips.

SpiNNaker \cite{spiNNaker2016}, proposed by the neuromorphic
engineering research group at Manchester university, is a
massively parallel, low power, neuromorphic computing board
designed to model very large, biologically realistic, spiking
neural networks in real time. The board consists of 65,536
identical 18-core processors, meaning  1,179,648 cores in total.
Each processor has an on-board router to form links with its
neighbours, as well as its own 128 MB of memory to hold synaptic
weights. Each core is an ARM968 chip manufactured using a 130 nm
process. The resultant chip is not a purely hardware neuromorphic
computer, but rather a massively parallel digital computer, with
over one million cores, which can simulate one thousand neurons
per core, leading to a total of one billion simulated neurons.

In 2014 Stanford University announced Neurogrid
\cite{Benjamin2014} structure. Neurogrid is a 6.5x7.5 square inch
circuit board composed of 16 custom designed chips, referred to as
NeuroCores. Each NeuroCore's analog circuitry covers 12x14 square
millimetres and is designed to emulate 65536 neurons, thus the
board can model a total of over one million neurons, and 6 billion
synapses implemented in a binary tree with links, with operates at
a total of 80 million spikes per second. The authors claim a
consumption of only 5 watt at full processing load.

Perhaps the most advanced neuromorphic chip is TrueNorth proposed
by IBM \cite{IBM2015}. TrueNorth is built in CMOS technology and
consists of 4096 hardware cores, each of them embedding 256
programmable hardware silicon neurons, which in turn have 256
programmable synapses. Over all the chip embeds over a million
neurons, with a total of over 268 million synapses, on a physical
CMOS wafer containing 5.4 billion transistors in 28 nm technology.
The authors claim an energy consumption of less than 100mW/chip,
meaning less then 20mW/$cm^2$, or 26pJ per synaptic event. They
also claim that the chip implements the first fully event driven
digital mixed synchronous-asynchronous neuromorphic architecture.
Applications of the chip in real tasks showed a consumption of
70mW while a recurrent network at biological real-time, and 65mW
in a multi-object detection and classification application, with
240x400-pixel 3-color video input at 30 frames-per-second.
TrueNorth is claimed to be the first neuromorphic implementation
that distances entirely from the classic processing-storage
paradigm, through the fact the neurons are silicon based not
emulated by conventional gates. Thus, the classic thinking of how
to program a computer also cannot be applied any more, since
concepts like memory access, data buses or CPU do not exist in
TrueNorth design. The authors designed for this reason an
end-to-end ecosystem complete with a custom simulator and a new
programming language for composing networks of neurosynaptic
cores, called corelet \cite{Amir2015} implemented in Matlab code.
With this, IBM also offers an integrated programming environment
with corelet libraries, conventional algorithms and applications
ported into the neuromorphic paradigm, and a teaching curriculum
called ``SyNAPSE University''.


\subsection{Human-Machine Interface Big Data Problems}

In our discussion of the sensor technologies to connect humans to
machines, we avoided some of the classic and data intensive
sensors such as different types of cameras and microphones. The
Psychophysiological sensors reviewed are not too data intensive.
The maximum EEG sampling rate was 20kHz. With a maximum of 256
sensors, we can have a 5mHz information flow from the human scalp.
ECG and other sensors are not worth discussing from a big data
perspective. Thus, it seems that the three most demanding data
sources are imagines, EEG and sound in order of the complexity of
data that can be acquired through these channels.

To process these data, clearly neuromorphic chips can offer a leap
forward in the ability of artificial AS to perform real-time
complex sensing and processing tasks. However, it is still too
early stage for this technology and is very unlikely that the
technology will be available in large scale and with a
general-consumer accessible strategy in the near future. As such,
we are back to the von Neuman and Harvard architectures, as the
most feasible processing to be used today. This also includes GPUs
and reconfigurable chips such as FPGAs.

From the 6 different Vs defining a big data problem, we comment in
what follows on how the human-machine interface problem is a big
data problem. The issue of volume arises specifically when dealing
with video, EEG and voice. However, the main big data challenges
within the context of this paper are more centred on the other Vs,
which are all apparent in this context. The timeseries nature of
most of these data, along with the chaotic nature of human mental
processing, makes the problem to be susceptible to high velocity
and high veracity. The multi-modality required to triangulate
different sources of data to understand human mental states make
the problem susceptible to high variety and variability features.
The use of these data to make informed trusting decisions and
facilitate human-machine interaction in TAS may depend on the
application and context, but in general we expect it to have clear
high value.

In summary, TAS bring big data challenges that need to be
considered when designing these systems.

 \section{Trusted Autonomous Systems: The Big Data Challenge Revealed}\label{TASBDCR}

This section concludes the paper with the big data open challenges
for trusted autonomy. In Sub-section~\ref{TheRoadToAutonomy}, we
will discuss autonomy to contextualise the discussion for the
reader. This is followed by bringing together the discussion on
Trust and the discussion on autonomy under the common objective of
this paper: trust-aware autonomous systems in
Sub-section~\ref{TrustAwareAutonomousSystems}.
Sub-section~\ref{OpenChallenges} offers some insight into some
open challenges and future research directions.

\begin{figure*}[!t]
 \centering
 \captionsetup{justification=centering}
 \includegraphics[width=12cm,height=6cm]{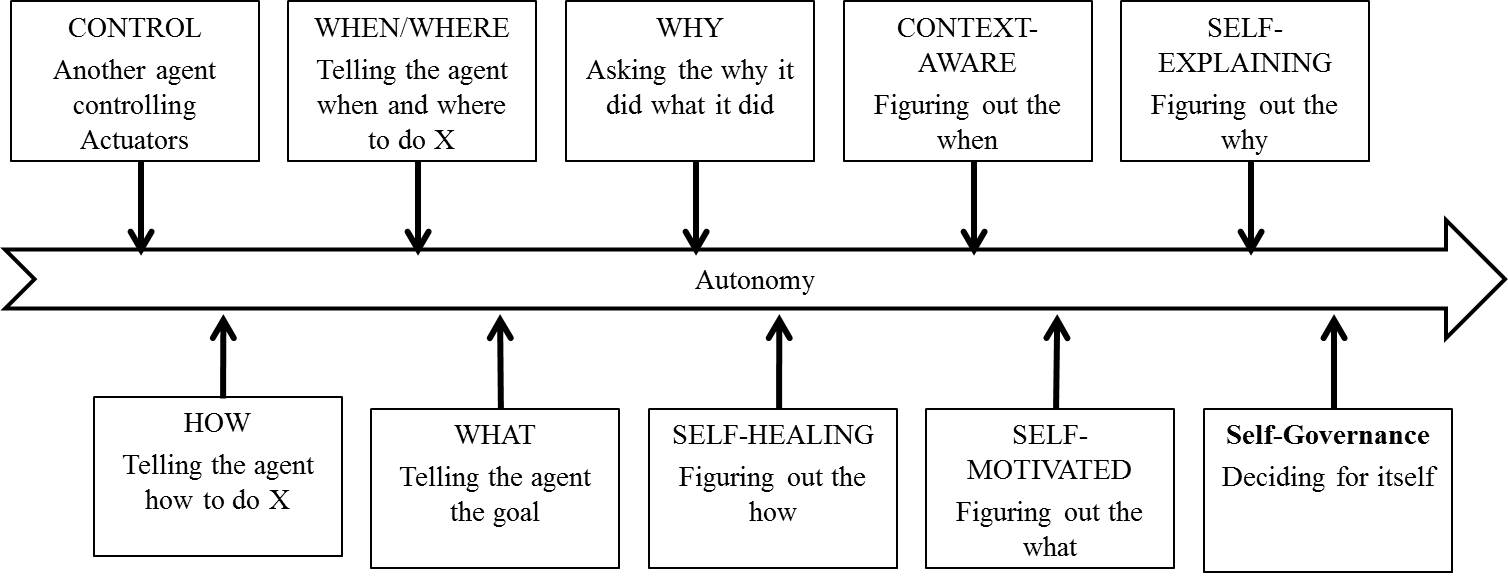}
 \caption{Autonomy Timeline.} \label{AutonomyTimeline}
 \end{figure*}

  \subsection{The Road to Autonomy}\label{TheRoadToAutonomy}

The evolution of autonomy in Figure~\ref{AutonomyTimeline} follows
similar principles to the evolution of trust. Autonomy is not a
binary concept. The concept exists on a wide scale with complete
absence of autonomy at the one end and full autonomy at the other.

We will explain autonomy using the WH-clauses: How (strategy),
When/Where/Whom (context), What (goal), and Why (rationality).
These questions are incremental in nature. Without the ability to
develop strategies, an actor can't manage and/or survive a
context. Without the ability to manage contexts, the actor can't
deal with goals. Without the ability to manage goals, the actor
can't explain and/or reason about its own behavior. The abilities
to develop strategy, manage context, manage goals, and explain and
reason can be hard-wired within an actor. When the actor is able
to develop these abilities by itself, it develops `self' and it
rises up in the hierarchy of autonomy. This ability to develop
`self' does not happen in vacuum. It is another set of algorithms
that work on a higher level of abstraction. Similar to learning to
learn, we still need to use learning techniques, but with an
appropriate representation, we can develop learning techniques
that are able to learn on their own.

The left hand side of the diagram starts with a zero-autonomy; a
dictatorship master-slave relationship, where zero autonomy is
equivalent to a slave with full obedience to its master. An
example we will use here to demonstrate the concept is in the case
of a bicycle. As a machine, it does not do anything unless an
external force (human or environment) exerts a force on it.

Let us add an electrical motor to the bicycle. The human may press
a button and it will start to move. The motor offers a control
without being aware of context. It will move in the same manner
independent of whether it is night or day, moving on a road or in
mud, and whether it is operated by an adult or a child. Once we
add sensors and increase the level of intelligence in the
controller, the machine starts to be `context-sensitive'. It will
behave differently in different contexts. We will follow a similar
philosophy to Abowd et.al.~\cite{abowd1999towards} to
differentiate between sensitivity to context; that is, the machine
is using context information, and what we will discuss later on as
`context aware'. For now, this machine is not `context aware' yet.

We can continue to improve the automation of the machine by adding
a planner and a GPS system to offer it an opportunity to decide on
goals. The human may ask the machine to go to a restaurant and the
machine will choose the restaurant, plan the route, and take the
human to a restaurant of its choice. The human may then need to
ask the `why' question; that is, why a certain route was chosen.
In this case, we need to use the planner within the machine or add
another layer of reasoning to offer the human an explanation of
the choices made by the machine.

While we do have a smart machine that automatically decides on a
route, adapts its control strategy to the information it is
sensing from the environment on terrain and driver
characteristics, is able to plan automatically, and reason about
its action, the level of autonomy of this machine is still very
limiting. What we have up to now is smart automation, with little
autonomy. Taking this level of intelligence to autonomy is the
vision of `autonomic computing' that was first introduced by Paul
Horn~\cite{horn2001autonomic} in his 2001 keynote address to the
National Academy of Engineers at Harvard University.

The ultimate aim of autonomic computing, or what is known now as
`self-X', is to have a self-governed system. This created a list
of characteristics to make this dream a reality including: the
system needs to have its own identity, be able to self-configure,
self-optimize, self-heal, self-protect, and become self-aware. We
will depart from some of these characteristics as we see the
evolutionary steps we will discuss for autonomy to subsume some of
them.

Autonomy starts to increase incrementally as we add more self-x
capabilities with the ultimate goal to achieve total
self-governing. The first feature is `self healing', the ability
to recognize problems (self-diagnosis) and recover automatically.
The system needs to be able to be equipped with the mechanisms for
it to discover its own strategies towards maintaining its health
level and functionality at the right level.

Moving in the hierarchy from context-sensitive to context aware,
the system needs to extend its ability to sense the environment
surrounding it to an ability where it senses the user's tasks,
user's intents, and know how to contribute to the optimization of
this task. This definition is consistent with Abowd
et.al.~\cite{abowd1999towards} who refer to a system as
context-aware ``if it uses context to provide relevant information
and/or services to the user, where relevancy depends on the user's
task."

A context aware system with the ability to do self-healing is
already too advanced in the autonomy hierarchy but still unable to
exercise full autonomy. It needs to have the ability to
autonomously generate its own goals through self-motivation and to
reason in its own way through self-reasoning. These features allow
the system to decide for itself what it needs to do and develop
its own rationale for the how, when, where and what.

With the above self-X capabilities, a system becomes fully
autonomous when it is able to integrate across these capabilities
to form a self-governed whole. Here, autonomy reaches its maximum
horizon.

 \subsection{Trust-Aware Autonomous
 Systems}\label{TrustAwareAutonomousSystems}

\begin{figure*}[!t]
 \centering
        \captionsetup{justification=centering}
        \includegraphics[width=15cm,height=14cm]{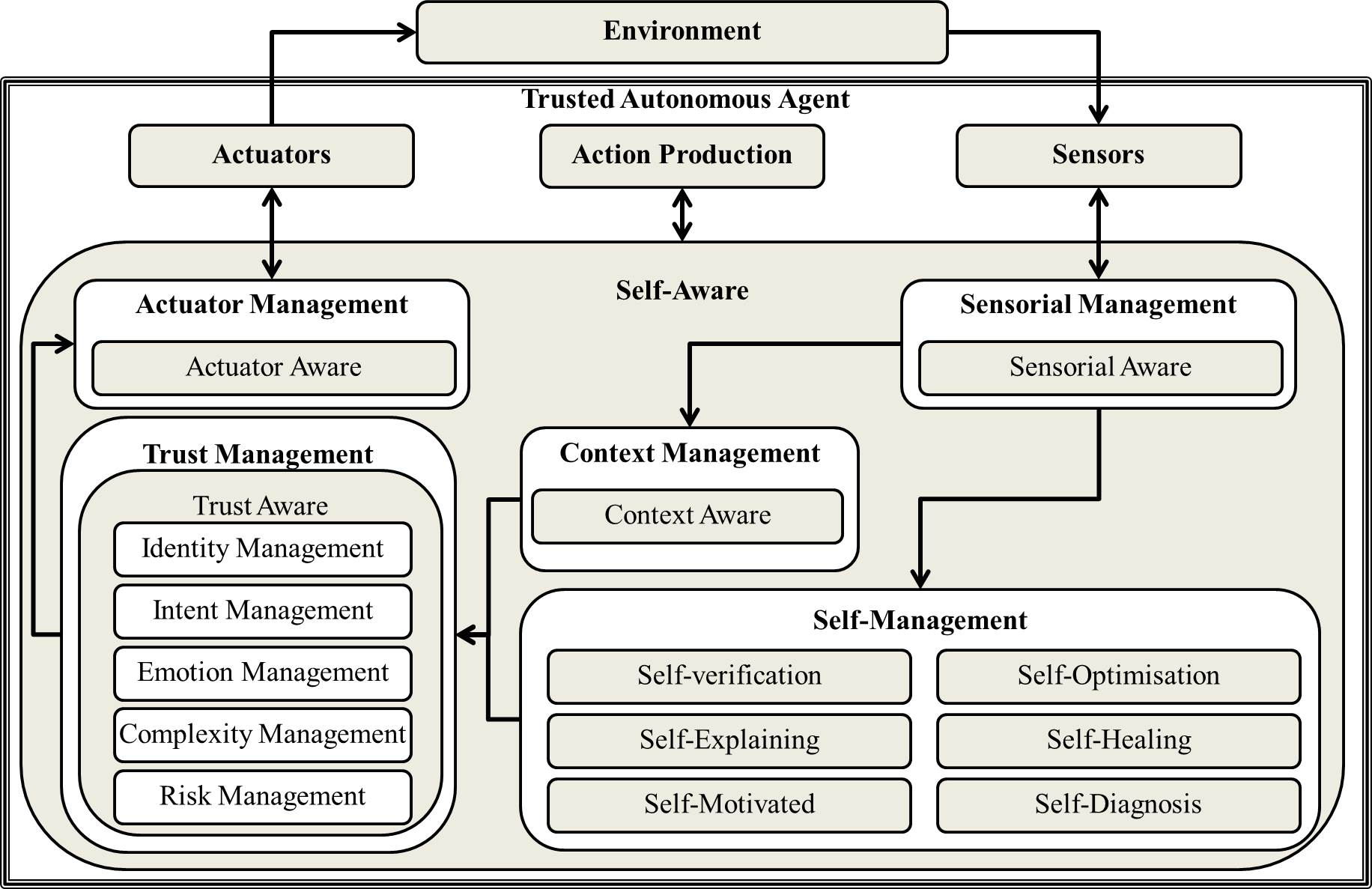}
        \caption{Trust-Aware Autonomous Systems.}
        \label{fig:TAS}

\end{figure*}

The road to trusted autonomy is not a simple addition of trust to
autonomy.

A trusted autonomous system (TAS) does not operate in a vacuum. It
interacts with other TASs and within an overall context. It needs
to take into account the complex interaction of the two concepts,
trust and autonomy, and the multi-actor environment they operate
within.

Each actor has its own objectives, goals, behavioral attributes,
capabilities, and resources. While these actors individually may
be trusted, the system as a whole may not be. The simple example
is counterintuitive and can better be explained with the
statement: if everyone does the right thing, it does not mean that
the group is doing the right thing. The challenge in this
statement lies within the boundary of the word `right': right for
whom and from whose perspective, and who has the authority to
define what is right? What is right from an individual actor
perspective may not be what is right from a system perspective. An
actor may wish to operate optimally (e.g. a robot would like to
reduce energy consumption to extend battery life), but the system
may need the task to be completed as quickly as possible, which
may result in a non-efficient use of the robot's battery. These
sorts of conflicting objectives between the individuals and the
group, the groups and the supergroups, and the supergroups and the
society create a level of interdependency that complicates the
concept of trust in social dynamics.

For a TAS to be trusted from a system-level perspective, it needs
to be self-aware of system-level objectives and work
collaboratively with other trusted-autonomous systems towards
achieving these objectives.

The dynamics associated with the interaction require continuous
sensing, processing, acting on, and verifying the environment it
is operating within and other autonomous systems, including
humans, in this environment.
Verification~\cite{zhang2015controller}, in particular, raises
more challenges and requires a significant amount of data in
real-time operations.

Figure~\ref{fig:TAS} attempts to capture the key building blocks
to form a TAS. The visual decomposition of these building blocks
is primarily for ease of understanding and is not meant to imply
that these building blocks can be designed or developed
independently. It is a conceptual functional decomposition of the
system, but it is not a proposal for a modular decomposition
because of complexities that are outside the scope of this paper.

It might seem counterintuitive to separate action production from
all modules that are focusing on self-management and
self-awareness. This is intentional conceptually because action
production mechanisms present a toolbox that any of the individual
building blocks can use and adopt when needed and as the situation
dictates. One can see action production as a service that gets
called upon by all other modules.

Secondly, we adopt a convention that self-management will require
self-awareness. We define self-awareness here as the system's
ability to map out and represent all its components virtually
({\it ie.} software wise), continuously measure their states, and
is able to identify if a component has been spoofed. This is a
stricter definition of self-awareness than some of the general
discussions of the concept in the literature. However, we take it
that a definition need to be self-encompassing, while systems can
exhibits different degrees of the concept being defined.

The following sub-section will highlight the big data challenges
for trusted autonomy.

 \subsection{Big Data Challenges for TAS Revealed}\label{OpenChallenges}

Trusted autonomy creates a series of big data challenges. We will
highlight below the six most prominent challenges that we have
identified.

\begin{enumerate}

 \item \textbf{Moving Sensor Network Challenge:}
An autonomous system is a moving sensor network that collects data
continuously, pro-actively, and in many instances intentionally to
improve its performance and achieve its mission's objectives.

We will put aside the resource-constrained environment of an
autonomous platform -- be it a hardware or a software agent, since
this will be the challenge we will discuss next. For now, we will
focus on the relationship between the agent's sensors and the
environment.

Many sensors will need to operate continuously. They can be
operating even when they are not used by the system. This is
particularly the case when a group of sensors are operating on the
same switch; thus, switching one of them off would simply switch
every sensor connected to the switch off. This raises an
attractive offer for the wider context that the autonomous agent
is operating within to leverage these data.

For example, while a Google car is moving, it may have a
temperature sensor to adjust the air conditioning system inside
the car. If the air conditioning system is switched off, this
temperature sensor can still be useful to transmit to the
intelligent transportation network localised temperature
information that can be integrated across the network to offer a
real-time environmental picture of the transportation network.

The above example imposes a big data challenge, when to allow for
this data to be transmitted and when to prohibit transmission? On
the one hand, transmitting the data benefits the wider system. On
the other hand, it overloads the car's battery as a result of the
extra energy required to transmit these data.

Moreover, it opens trust challenges. The local temperature of the
car may get seen by some as a not-threatening privacy data and get
transmitted to third parties. A situation that can create security
and privacy hazards, whereby the localised temperature can be used
as a heat-signature for tracking the car by an entity that does
have a privilege to do so and may not be allowed to recieve GPS
information from the car.

A second factor to consider in this challenge is related to the
words `pro-actively and intentionally'. The autonomous agent needs
to decide which data is useful for its own context. In a dynamic
environment where the situation is changing continuously, to
decide which data is more useful to get, the autonomous entity
needs to continuously be listening to all its sensors, processing
the information and making a decision on the `value' of the data
so that it is able to pro-actively decide which sensor needs to be
targeted for more data and to intentionally decide which decision
is a higher priority at this point of time and the data needed for
this decision.

The problem above is very complex. To allow the system to
continuously monitor all of its sensors, the system may need to
reduce the sample rate for those sensors that may not be important
at one point of time, then dynamically adjust the sampling rate as
they become important. This is a very challenging problem.

To automatically adapt the data acquisition strategy in a sensor
network raises all sorts of big data problems, from the fact that
the data arrives with different level of noise, the environment is
dynamic and concepts change over time, to issues of sensor
reliability and the trust-level associated with each sensor.

 \item \textbf{Load-Balancing and Value Challenge:}
Almost all TASs would have physical constraints in terms of memory
and processing capacity. This resource-constrained environment
compounds the complexity of too much available data, and the too
little knowledge that we can draw out of it.

A TAS, similar to a human, encounters data as it is operating that
may not be useful for itself, but for others in the environment.
Imagine two robots, one is collecting tomatoes and the other is
collecting weeds. The robot collecting the tomatoes may sense the
existence of weeds. Whether or not it should send this information
to the other robot or not would depend on many factors. However,
even the decision itself raises the question of how to balance the
load on its memory, processing capacity, and battery life in these
situations?

The above question is a classic question in autonomous systems,
but big data makes the question even more demanding. The data the
TAS collects, even unintentionally, may hide a great deal of
values to the system. The amount of processing and resources to be
allocated for this task necessitate a level of processing that can
threaten the survivability of the system in a harsh operating
environment. Nevertheless, ignoring this information can threaten
even more the survivability of this system.

 \item \textbf{Trust-Awareness Cognitive Challenge:}
For an autonomous system to be trusted, it can't focus purely on
its primary mission as it needs to make every interaction it is
involved within as part of its mission. This increases the data
load on the system and creates higher than ever demands on the
system to process these data in real-time.

The primary challenge here is related to velocity, veracity and
value. An agent that is acting in a real-time environment needs to
mine the data it is receiving very quickly so that its actions
take into account any patterns that have arisen in the moments
before actions are made. A real-time situation is dynamic by
nature. The changing situation needs to integrate its current
states with new information. This integration process can be very
expensive. On the one level it is a classic information fusion and
belief update problem. On another level, it can trigger the
activation of almost every component of a self-governing system.
For example, the system may realise it needs to self-optimize in
response to a self-diagnosed failure in one of its sensors, while
simultaneously the self-healing module is attempting to recover
the data through self-organising the rest of the sensor network.
The synchronization of process and action in real-time is one of
the grand challenges of TAS.

 \item \textbf{Trusted Swarm Challenge:}
A TAS does not work in isolation of other autonomous systems in
the environment. Trust offer a path for autonomous systems to
manage the complexities discussed above in the same way trust
reduces complexity in a social system. It allows for delegation as
a mechanism to manage complexity and distribute load among the
system's components.

In the previous challenge, we discussed an example which involved
two types of complexity that the agent is faced with. One is
related to the amount of processing an agent needs to do to
synchronise its action production with a fast interaction. The
second is related to loss of information as a result of sensor
failure or a conscious decision by the agent. In any of these
situations, if the autonomous system trusts another autonomous
system, the former can use the latter to overcome some of these
challenges, assuming that the latter can do the job.

The above example focuses on a two-entity interaction, but the
true benefits from trust becomes clearer when it happens within a
swarm, whereby a whole lot of delegations can occur in the system
among the trusted autonomous agents so that overall mission
success is maximized. An example of this would be apparent in the
mining industry where a swarm of machines need to cooperate to
maximize the probability of discovering a suitable place to mine.

The degree of coordination and delegation that occur in this
example dramatically increase the data flow among the agents,
whether these data are communicated directly through direct
communications channels or indirectly through behavioural changes
to signal a need for help. Again, this problem raises big data
challenges because of value, velocity and veracity. There is a
significant value add for the agents to cooperate in situations
such as those described above to maximize the overall mission
success of the swarm. These interactions are not free of
complexity as the situation can be changing fast introducing a
great deal of velocity challenges, and the fast sampling rate can
be an extreme source of veracity which can cause instability in
any inferring mechanism used in the system.

 \item \textbf{Trusted Cyborg Challenge:}
An artificial TAS -- such as a software agent or a robot -- is
expected to work in harmony with humans. With appropriate use of
advanced AI techniques, the artificial and human entities need to
work as a single entity, an interdependency that is known as
symbiosis.

The challenges arises on how to make the silicon and biological
agents work together as one; that is, establishing a truly
symbiotic relationship to allow the machine and the human to meld
as one entity. The coupling of the two creates a new type of
agents, a Cyborg (human-machine coupled computational decision
making engine).

The machine needs to understand the human. However, the complexity
arising from the diversity of human attitudes and personalities is
huge. One possible mechanism to manage this complexity is to
directly obtain objective data on the human to manage the
ambiguity that comes with human communication. The data needs to
be collected through mechanisms that do not interfere with the
performance of the human on the job. The use of body sensors and
passive external sensors such as cameras and microphones can offer
an unprecedented amount of data to allow this integration to
occur.

The Cyborg's brain needs to have efficient interfaces between the
biological and silicon brains so that both human and machine work
seamlessly together as a single entity. We are not suggesting here
to connect human brains to silicon or any type of science fiction.
The suggestion is to integrate the decision making between the two
in the same way two close friends work together to solve a
problem.

The Cyborg's brain needs on-board data processing. The real-time
and volume of the data necessitate a very efficient use of the
hardware. Technologies similar to those discussed in
Section~\ref{BDHMI} such as neuromorphic computing, FPGA and other
implementations of neural networks on chips can be very
attractive. Equally important here is the need for real-time
signal processing techniques capable of sensing the context to
improve signal cleaning, and adapt sampling rate when and as
needed, with appropriate dynamic calibration models that can fuse
different sources of data from the human and dynamically adapt the
parameters and structure of the model in response to changes in
the environment.

 \item \textbf{Trusted Cyborg Swarm Challenge:}
A group of Cyborgs make a Cyborg swarm. The complexity of a
network of Cyborg swarms, an extension of a problem we called in
our previous research as Cognitive Cyber Symbiosis
(CoCyS)~\cite{abbass2015trusted,abbasscrt1}, needs to be managed
to allow this decision network of Cyborgs to operate efficiently.
CoCyS is a decision network of humans and machines. A Cyborg swarm
is a network where each node is a Cyborg's brain.

A Cyborg swarm is the next generation swarm intelligence system,
where each human can look after a swarm of artificial TASs, and
the Cyborg population works together as a swarm to solve far
reaching complex.

Take for example a situation like a post-disaster recovery, where
a tsunami has destroyed a village. Each site may have a Cyborg
consisting of a human and a swarm of artificial TASs. The Cyborg
network needs to coordinate activities, share information, and
work as one team to assist the village in recovery. Such a network
should recover from the disaster a few orders of magnitude faster
than what we do today, as a result of the close integration of
human and machine.

The challenges of forming such a network define scientific
frontiers of this century. On the one hand, we have computational
challenges to design appropriate AI systems that can handle this
level of distributed control and complex and dynamic context. On
the other hand, the amount of data in a Cyborg swarm is huge.
These data are not just useful for the swarm to make decisions for
recovery. They are also useful for future disaster situations,
where it is important that these data get collected from the swarm
as well.

\end{enumerate}

Putting the six challenges above together, theoretical and
practical aspects of the big data challenges for trusted autonomy
draw a roadmap for the research needed to overcome these problems
in practice.

%


\end{document}